\documentclass[pdflatex,sn-nature]{sn-jnl}

\usepackage{graphicx}%
\usepackage{multirow}%
\usepackage{amsmath,amssymb,amsfonts}%
\usepackage{amsthm}%
\usepackage{mathrsfs}%
\usepackage[title]{appendix}%
\usepackage{xcolor}%
\usepackage{textcomp}%
\usepackage{manyfoot}%
\usepackage{booktabs}%
\usepackage{algorithm}%
\usepackage{algorithmicx}%
\usepackage{algpseudocode}%
\usepackage{listings}%
\usepackage{soul}%
\usepackage{overpic}
\usepackage{lineno}

\begin{document}

\title[Article Title]{Performance-driven Computational Design of Multi-terminal Compositionally Graded Alloy Structures using Graphs}


\author*[1]{\fnm{Marshall D.} \sur{Allen}}\email{marshdallen10@gmail.com}

\author[2]{\fnm{Vahid} \sur{Attari}}\email{attari.v@tamu.edu}

\author[2]{\fnm{Brent} \sur{Vela}}\email{brentvela@tamu.edu}

\author[2]{\fnm{James} \sur{Hanagan}}\email{james\_hanagan@tamu.edu}

\author*[1]{\fnm{Richard} \sur{Malak}}\email{rmalak@tamu.edu}

\author*[1,2]{\fnm{Raymundo} \sur{Arr{\'o}yave}}\email{raymundo.arroyave@tamu.edu}

\affil[1]{\orgdiv{J. Mike Walker '66 Department of Mechanical Engineering}, \orgname{Texas A\&M University}, \orgaddress{\city{College Station}, \postcode{77843}, \state{TX}, \country{USA}}}

\affil[2]{\orgdiv{Department of Materials Science and Engineering}, \orgname{Texas A\&M University}, \orgaddress{\city{College Station}, \postcode{77843}, \state{TX}, \country{USA}}}


\abstract{The spatial control of material placement afforded by metal additive manufacturing (AM) has enabled significant progress in the development and implementation of compositionally graded alloys (CGAs) for spatial property variation in monolithic structures. However, cracking and brittle phase formation have hindered CGA development, with limited research extending beyond materials design to structural design. Notably, the high-dimensional alloy design space (systems with more than three active elements) remains poorly understood, specifically for CGAs. As a result, many prior efforts take a trial-and-error approach. Additionally, current structural design methods are inadequate for joining dissimilar alloys. In light of these challenges, recent work in graph information modeling and design automation has enabled topological partitioning and analysis of the alloy design space, automated design of multi-terminal CGAs, and automated conformal mapping of CGAs onto corresponding structural geometries. In comparison, prior gradient design approaches are limited to two-terminal CGAs. Here, we integrate these recent advancements, demonstrating a unified performance-driven CGA design approach on a gas turbine blade with broader application to other material systems and engineering structures.}


\maketitle

\section{Introduction}\label{sec:intro}
The traditional approach of using a single material to fabricate critical components is becoming increasingly challenging. Prioritizing the selection or design materials that provide the best overall performance across multiple criteria overlooks the fact that performance requirements are typically localized. This leads to materials being over-specified in certain regions of the part while simultaneously failing to meet minimal performance standards in others, resulting in considerable design inefficiencies. As technologies advance and performance demands grow more complex, the quest for materials with universally optimal properties across all localized requirements has become increasingly challenging and, in many cases, fundamentally impractical. 

In this paper, we demonstrate the systematic design of multi-terminal compositionally graded alloys (CGAs) based on spatial performance-based materials design criteria from a corresponding 3D structure. CGAs are a type of functionally graded material (FGM) where the spatial variation of material composition is used to control the spatial variation of material properties. Metal additive manufacturing (AM) techniques such as laser-powder directed energy deposition (LP-DED) \cite{banerjeeMicrostructuralEvolutionLaser2003, collinsLaserDepositionCompositionally2003, bobbioAdditiveManufacturingFunctionally2017c, bobbioCharacterizationFunctionallyGraded2018}, wire arc additive manufacturing (WAAM) \cite{wangCharacterizationWireArc2018, rodriguesSteelcopperFunctionallyGraded2022, johnStudiesSituAlloy2024}, and powder bed fusion (PBF) \cite{walkerMultimaterialLaserPowder2022a, weiCompositionallyGradedAlxCoCrFeNi2022, wuReviewExperimentallyObserved2023} are often used for manufacturing CGAs due to their capability to control alloy composition spatially by altering the ratio of different alloy feedstocks during fabrication. Despite manufacturing efforts, CGAs have not been implemented broadly due to challenges with materials design and manufacturing.

The primary materials design challenge for CGAs is avoiding the formation of deleterious brittle phases at intermediate compositions \cite{reichardtAdvancesAdditiveManufacturing2021}. To address this challenge, Hofmann et al. proposed the use of ternary phase diagrams for CGA design, which can be modeled using the CALculation of PHAse Diagrams (CALPHAD) approach and visualized in 2D \cite{hofmannCompositionallyGradedMetals2014}. These phase diagrams can be used as maps to visually discern if a feasible gradient between two terminal (\lq{endpoint}\rq) alloys exists and, if so, to manually design a manufacturable CGA path between these alloys. This approach has been further augmented with Scheil-Gulliver simulations \cite{BOCKLUND2020100689, bocklund2021computational, bobbioDesignAdditivelyManufactured2022b} and cracking susceptibility metrics \cite{yangDesignMethodologyFunctionally2023a}. This augmented methodology has recently been released in the open-source software MaterialsMap \cite{SUN2024102153}. 

Meanwhile, Kirk et al. introduced a computational methodology \cite{kirkComputationalDesignGradient2018, kirkComputationalDesignCompositionally2020d} based on a robotic motion planning algorithm, RRT*FN \cite{adiyatovRapidlyexploringRandomTree2013a, karamanSamplingbasedAlgorithmsOptimal2011}. This methodology used the CALPHAD method to perform equilibrium thermodynamic calculations to train a binary machine learning classifier to distinguish between compositions expected to form deleterious phases vs. manufacturable phases. The trained binary classifier was then used as the obstacle model for RRT*FN; the RRT*FN algorithm would then operate automatically, in any number of dimensions, optimizing the gradient path for a given objective function while avoiding the formation of deleterious phases. Several experimental demonstrations were performed based on this methodology \cite{eliseevaFunctionallyGradedMaterials2019a, kirkComputationalDesignCompositionally2020a}. This approach was later augmented with a sampling procedure that considers all subspaces of the alloy design space in the gradient design process \cite{allenSubspaceInclusiveSamplingMethod2021}. Most recently, on-the-fly CALPHAD sampling was proposed to reduce the overall computational cost of this approach \cite{PriceNovelPathFramework2024}. 

\begin{figure}
    \centering
    \includegraphics[width=\columnwidth]{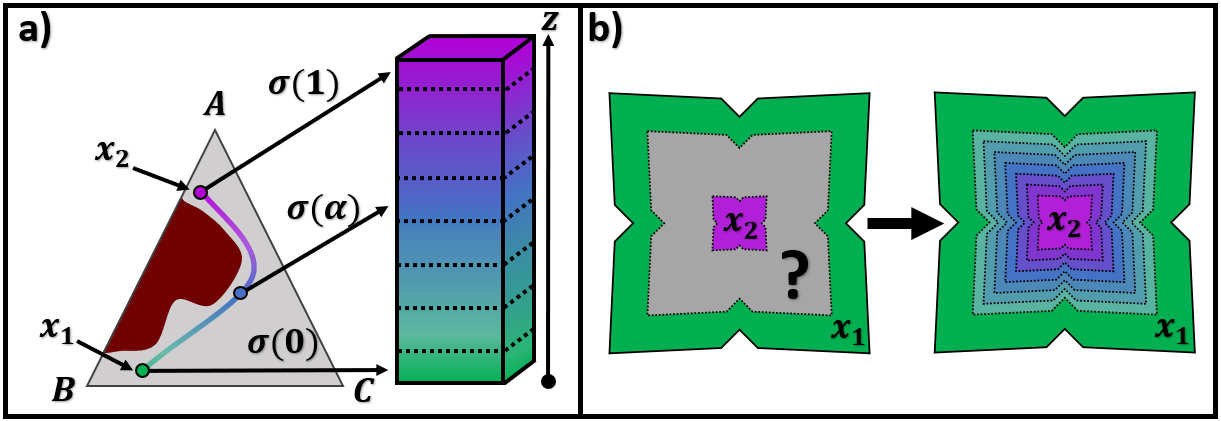}
    \caption{\textbf{Illustrative example of mapping a two-terminal CGA from a materials domain to the physical domain.} \textbf{a)} shows the design of a one-dimensional compositionally graded alloy path, $\sigma$, between terminal alloy compositions \( x_{1} \) and \( x_{2} \) in a synthetic ternary alloy system consisting of elements $\textbf{A}$, $\textbf{B}$, and $\textbf{C}$, with the maroon region indicating deleterious compositions. This compositional gradient can be easily mapped into physical space by correlating compositions along the gradient, $\sigma(\alpha), \alpha \in [0,1]$, to a corresponding physical coordinate, such as build height $z$. Similarly, \textbf{b)} demonstrates how a one-dimensional gradient can be applied to a structure with arbitrary geometry by segmenting it into encompassing shells or layers and performing a similar mapping.}
    \label{fig:one-dimensional-gradient-example}
\end{figure}

These methods have successfully identified two-terminal CGAs in multi-dimensional alloy spaces, with RRT*FN offering dimensional scalability. A two-terminal CGA path is one-dimensional, represented as a continuous function $\sigma : [0,1] \rightarrow (\mathbb{R}^n)$. The composition along the path is described by a single index variable $\alpha \in [0,1]$, which scales with the Euclidean distance $l$ along the path in $(\mathbb{R}^n)$, from one terminal alloy ($\alpha = 0$) to another ($\alpha = 1$). Figure~\ref{fig:one-dimensional-gradient-example}a illustrates this for a CGA between terminal alloys $x_{1}$ and $x_{2}$, where the path index $\alpha$ can also map to a physical coordinate, such as the build direction $z$. Figure~\ref{fig:one-dimensional-gradient-example}b extends this concept to arbitrary structures based on terminal alloy placement.

While these methods enable CGA design, they face notable limitations. Visual approaches are restricted to ternary systems and often yield suboptimal paths compared to automated algorithms. Robotic path planning, though capable of exploring high-dimensional spaces, provides limited insight into the feasible design space. Designers cannot determine whether a manufacturable path exists between terminal alloys without running the RRT*FN algorithm, and failure to generate a path after numerous iterations leaves ambiguity about the absence of a solution or the need for more iterations. Although the RRT* algorithm is probabilistically complete \cite{karamanSamplingbasedAlgorithmsOptimal2011}, the iterations required to find a path vary significantly with the state space, with narrow feasible regions often requiring far more computational effort.

Another limitation is the material-biased design process, which typically begins by selecting intermediate compositions to form a defect-free joint between two specific alloys. This approach unnecessarily constrains the design space when the goal is to achieve specific properties irrespective of composition, assuming that a defect-free gradient will always satisfy design requirements—an assumption that may not hold. Removing this bias allows for a systems approach, as described in G. B. Olson's seminal work \cite{olsonComputationalDesign}, linking desired properties and structural performance back to structure and processing. However, specific alloys may still need to be joined in cases dictated by material availability, standards, or established applications.

The most \emph{fundamental limitation} of the current approaches to CGA design is the confinement of existing methods to two-terminal problems, restricting gradients to connect only two terminal alloys. This precludes solutions requiring gradients involving more than two alloys, limiting their ability to address complex design challenges.

Due to these limitations, current authors M. Allen, R. Malak, and R. Arr{\'o}yave previously proposed several advancements for formal data modeling and automated design of CGAs and corresponding structures---see \emph{Methods} section for more details. First, they proposed the Alloy Topology-Linked informAtion Schema (ATLAS), a formal hierarchical data model for representing the alloy chemistry design space as a labeled property graph (LPG), as described in Allen \emph{et al.} \cite{allenGraphDatabaseSchema2024, allenSystematicDesignCompositionally2025}. In this framework, the \lq{ATLAS materials graph}\rq~captures the topology of the alloy composition space, where nodes represent unique alloy compositions, and edges indicate compositionally similar alloys that could potentially be joined in a CGA. 

Building on ATLAS, Allen \emph{et al.} proposed a general multi-terminal formulation of the CGA design problem \cite{allenGraphAlgorithmDesign2024, allenSystematicDesignCompositionally2025} as a minimum Steiner tree problem in graphs (STP) \cite{dreyfusSteinerProblemGraphs1971}, generalizing the concept of shortest path optimization to \emph{arbitrarily many} terminal alloy compositions. The tree given by solving the STP for a given set of terminal alloys represents a multi-terminal CGA for those alloys, $\tau_{\text{CGA}}$. 

Lastly, Allen \emph{et al.}  introduced the Tree-based Material Adjacency Propagation (TreeMAP) algorithm for conformally mapping an arbitrary multi-terminal gradient from a graph tree, $\tau_{\text{CGA}}$, onto a corresponding physical part \cite{allenGraphAlgorithmDesign2024, allenSystematicDesignCompositionally2025}. Before this work, compositional variation of CGAs was typically considered with respect to placing a two-terminal CGA along a single Cartesian, cylindrical, or spherical axis in manufacturing \cite{hofmannCompositionallyGradedMetals2014, walkerMultimaterialLaserPowder2022a}. To the authors’ knowledge, neither identifying multi-terminal CGAs nor embedding them in complex structural designs had been achieved prior to their introduction of the STP formulation and TreeMAP algorithm.

In this work, we demonstrate the combined capabilities of the ATLAS framework \cite{allenGraphDatabaseSchema2024, allenSystematicDesignCompositionally2025} for alloy informatics and the TreeMAP algorithm \cite{allenGraphAlgorithmDesign2024, allenSystematicDesignCompositionally2025} for conformal gradient material placement in the performance-driven computational design of a gas turbine blade with three terminal alloys. The materials design process begins with a simplicial grid sampling of compositions in the Cr-Nb-V-W-Zr quinary system and its subsystems, using a 0.05 mole fraction spacing. This refractory system was selected since alloys in these systems have been generally shown to form body-centered-cubic (BCC) solid solutions with desirable high-temperature properties \cite{senkovDevelopmentRefractorySuperalloy2016, senkovDevelopmentExplorationRefractory2018}. A deep learning regression model~\cite{attari2024decoding} predicts relevant properties, imputing missing data, correcting errors, and calculating a property-based cost function. Additionally, CALPHAD equilibrium calculations and Scheil-Gulliver solidification simulations \cite{Scheil+1942+70+72, gulliver1913quantitative} are used to identify compositions expected to form a single BCC phase.

Using ATLAS, we construct a materials graph from this dataset, where nodes represent unique alloy compositions and edges indicate compositionally feasible connections. Nodes are filtered based on phase and property requirements, and the resultant materials graph is partitioned into constrained subgraphs. Further filtering based on design criteria yields a single subgraph from which we identify three terminal alloys optimized for localized performance objectives relevant to the turbine blade design problem. These terminal alloys exhibit specialized properties \emph{unattainable by any single alloy} in the dataset. From the selected terminal alloys, we design a multi-terminal CGA using a Steiner tree heuristic solver \cite{mehlhornFasterApproximationAlgorithm1988a} and a cost function to minimize worst-case cracking susceptibility, computed using the deep learning regression model. The resulting CGA connects the terminal alloys through feasible compositions that exhibit low cracking susceptibility. The CGA is then applied to a discretized 3D gas turbine blade model, with two of the three terminal alloy regions spatially defined. Finally, we use the TreeMAP algorithm to propagate the compositional gradient through the 3D structure. TreeMAP efficiently maps the gradient by traversing the discretized geometry with an intermediate graph representation, ensuring material adjacency requirements are satisfied and infeasible compositions are avoided. The final turbine blade design integrates terminal alloy regions tailored for localized performance conditions, seamlessly joined by an optimized compositional gradient in conformal 3D layers.

\section{Results}\label{sec:results} 
\subsection{Terminal Alloy Design Problems} \label{sec:design_problem}

\begin{figure}
\centering
\includegraphics[width=0.9\textwidth]{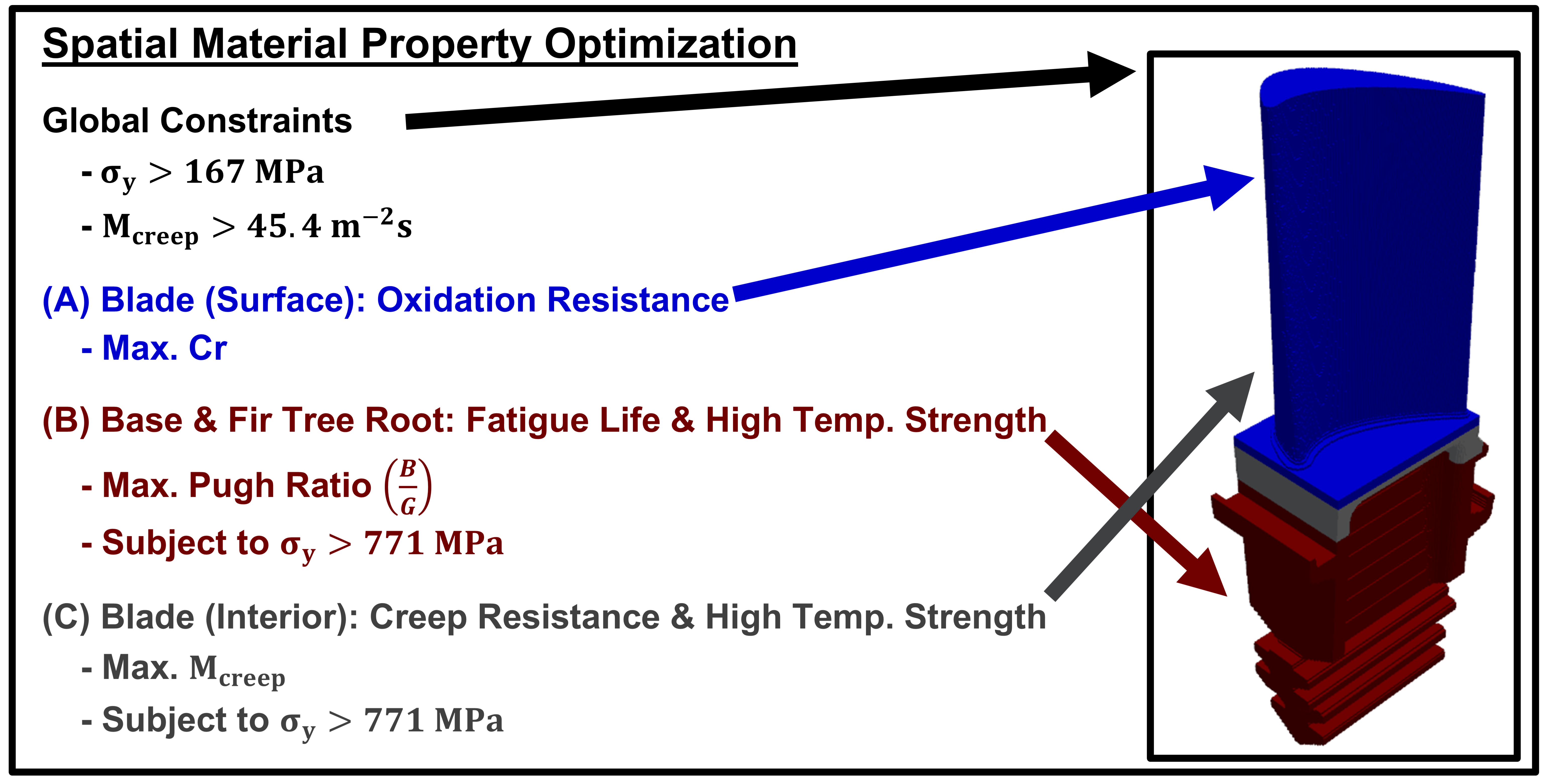}
\caption{\textbf{Localized spatial property optimization objectives \& constraints for the gas turbine blade design demonstration.}}\label{fig:design_problem_definition}
\end{figure}

The spatial material property/performance optimization problems for this demonstration are illustrated in Fig.~\ref{fig:design_problem_definition}. We first set an overarching global requirement that the high-temperature yield strength, $\sigma_\text{y}$, and creep merit index \cite{REED20095898}, ${M_{\text{creep}}}$, should be above the 25th percentile of the distribution in order to avoid the use of extremely poor materials in the structure. For $\sigma_\text{y}$, we assumed $1000^\circ$C was a representative operating temperature to evaluate the yield strength for gas turbine blade alloys \cite{REZAZADEHREYHANI2013148}. This results in respective lower bounds of 167 MPa for $\sigma_\text{y}$ and 45.4 $\text{m}^{-2}\text{s}$ for ${M_{\text{creep}}}$. We expressed the optimization of terminal alloys for the compositional gradient as three separate optimization problems to identify terminal alloy nodes $x_{A}$, $x_{B}$, and $x_{C}$ in the discrete nodal $Alloy$ compositions $V_A$ in the feasibility-constrained ATLAS materials graph. Note that all property percentile thresholds were calculated based on the properties of the feasible single-phase BCC compositions, not the entire dataset. This avoided the selection of nominally optimal but ultimately infeasible terminal alloys.

For the first terminal alloy, $x_{\text{A}}$, we wanted to maximize the oxidation resistance performance, as indicated by Cr content, on the high-temperature flow-facing surfaces of the blade. We assumed higher Cr content approximately indicated higher oxidation resistance, as the likely formation of stable passivating Cr$_2$O$_3$ layers increases with Cr chemical activity, which in turn is higher at high Cr concentrations in the alloy. We recognize that there are more rigorous metrics for characterizing oxidation resistance in alloys \cite{SaucedaOxidationLandscape2022}. However, implementing these metrics would not change the central methodology demonstrated in this work and, thus, was considered out of scope. This optimization problem for the first terminal alloy can be expressed as

\begin{align}
\max_{v_A \in V_A} \ & \text{Cr}(v_A) \notag \\
\text{subject to:} \ & \sigma_\text{y}(v_A) > 167 \, \text{MPa}, \\
& M_{\text{creep}}(v_A) > 45.43 \, \text{m}^{-2}\text{s} \notag
\end{align}
\label{eq:blade_surface_optimization}

\noindent where $V_A$ is the set of all $Alloy$ nodes in the ATLAS materials graph. 

Next, for the second terminal alloy, $x_{\text{B}}$, with high-temperature fatigue life and strength in the base \cite{houInvestigationFatigueFailures2002} (including the fir tree root), we wanted to maximize the ductility (indicated by Pugh ratio \cite{pughXCIIRelationsElastic1954}) as a metric for improved fatigue life, while achieving $>$95th percentile $\sigma_\text{y}$. This optimization problem can be expressed similarly as follows:

\begin{align}
\max_{v_A \in V_A} \ & \frac{\text{B}(v_A)}{\text{S}(v_A)} \notag \\
\text{subject to:} \ & \sigma_\text{y}(v_A) > 771 \, \text{MPa}, \\
& M_{\text{creep}}(v_A) > 45.43 \, \text{m}^{-2}\text{s} \notag
\end{align}
\label{eq:fir_tree_root_optimization}

\noindent where $\frac{\text{B}(v_A)}{\text{S}(v_A)}$ is the Pugh ratio (bulk modulus, $B$, divided by shear modulus, $S$) of a given nodal alloy composition, $v_A$.

Lastly, while not geometrically fixed to a specific location, the last spatial material optimization was for a third terminal alloy, $x_{\text{C}}$, with high-temperature creep resistance and strength in the blade interior \cite{REZAZADEHREYHANI2013148}. For this alloy, we wanted to maximize ${M_{\text{creep}}}$ with $>$95th percentile $\sigma_\text{y}$. This optimization problem can be expressed as

\begin{align}
\max_{v_A \in V_{A}} \ & M_{\text{creep}}(v_A) \notag \\
\text{subject to:} \ & \sigma_\text{y}(v_A) > 771 \, \text{MPa}, \\
& M_{\text{creep}}(v_A) > 45.43 \, \text{m}^{-2}\text{s} \notag
\end{align}
\label{eq:blade_interior_optimization}

\noindent Here, $x_{\text{C}}$ is not fixed to a specific geometric region within the part since its position depends on the number of compositions in the gradient to join the three terminal alloys in this design. This consideration is a direct and important consequence of integrating physical design, manufacturing considerations, and materials design into a unified methodology.

\subsubsection{Terminal Alloy Design}
First, three terminal alloy compositions were identified from the only constrained subgraph of the ATLAS materials graph for this Cr-Nb-V-W-Zr quinary system, $G_{\alpha} \in G_{\text{AM}}$ that surpassed the constraints for the spatial property optimization problems. For alloy $x_{A}$, an oxidation-resistant alloy with maximum Cr content, the optimal composition was identified as Cr\textsubscript{45}V\textsubscript{35}W\textsubscript{20}. This alloy has more than twice the Cr content of many Ni-based superalloys. For alloy $x_{B}$, an alloy with maximum fatigue-resistance (indicated by Pugh ratio) and $\sigma_\text{y} \geq 771~\text{MPa}$, the optimal composition was identified as Cr\textsubscript{15}Nb\textsubscript{25}V\textsubscript{40}W\textsubscript{20}. This alloy has a predicted Pugh ratio of 2.71, above the typical 1.75 ductility threshold, and predicted $\sigma_\text{y} = 830~\text{MPa}$. Lastly, for alloy $x_{C}$, an alloy with maximum high temperature creep-resistance and $\sigma_\text{y} \geq 771~\text{MPa}$, the optimal composition was identified as Cr\textsubscript{30}Nb\textsubscript{5}V\textsubscript{40}W\textsubscript{25}. This alloy has a predicted $M_{\mathrm{creep}} = $ 429 m\textsuperscript{-2}s, and $\sigma_\text{y} = 786~\text{MPa}$. With these terminal alloy compositions selected, the next step is to proceed with identifying a multi-terminal CGA.

\subsubsection{Multi-terminal Gradient Alloy Design}

\begin{figure}
\centering
\includegraphics[width=0.8\textwidth]{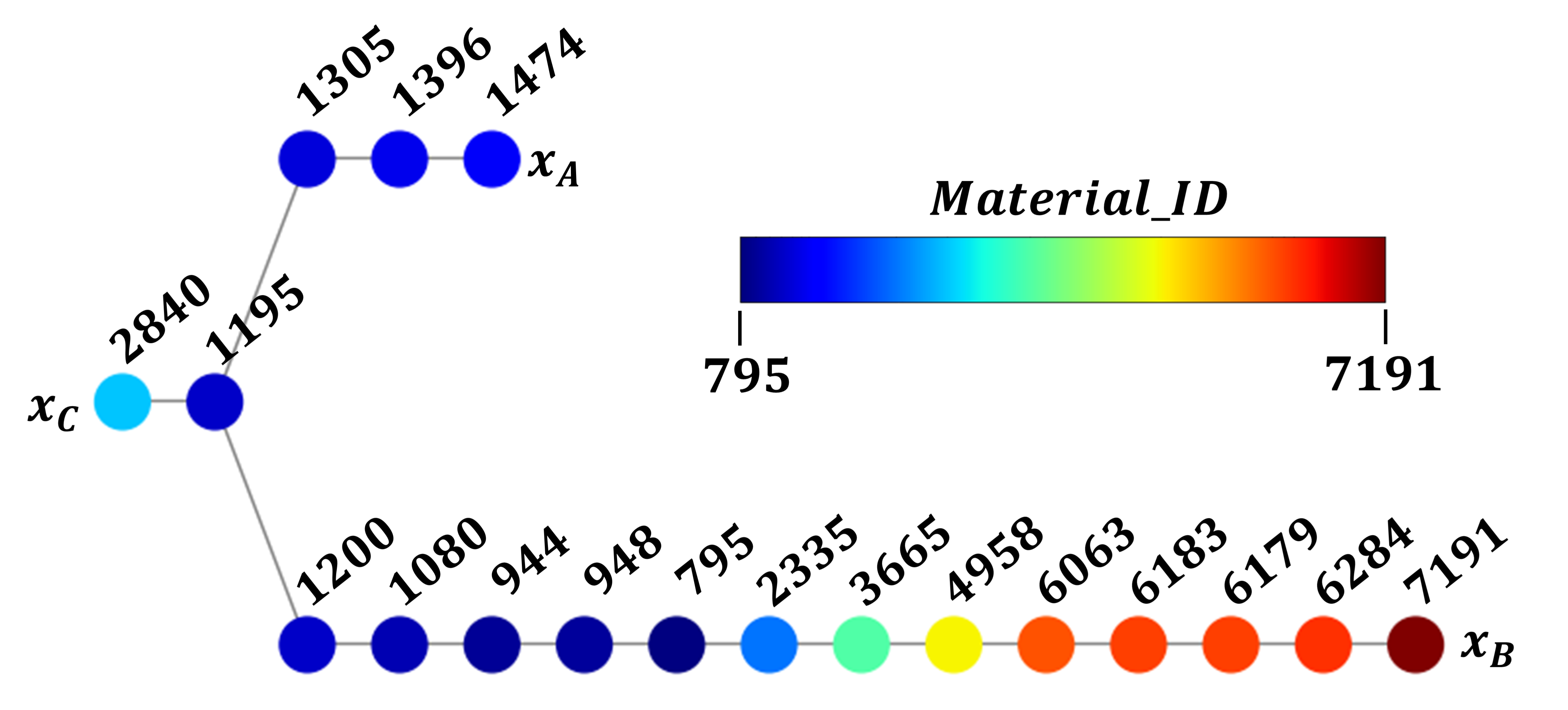}
\caption{\textbf{The optimized CGA tree, $\tau_{\text{CGA}}$, with nodal $Alloy$ compositions and $JOINS\_TO$ edges between the three terminal alloys $x_{A}$, $x_{B}$, \& $x_{C}$, labeled and colored by $Material\_ID$}}\label{fig:gradient_tree}
\end{figure}

With these terminal alloys, we then used a cost function to minimize the worst-case cracking susceptibility and an existing heuristic solver \cite{mehlhornFasterApproximationAlgorithm1988a} for the STP. From this input, we designed a multi-terminal gradient tree, $\tau_{\text{CGA}}$, that connects the three terminal alloys. The topological structure of the tree is visualized in Fig.~\ref{fig:gradient_tree} alongside a unique nodal $Alloy$ composition identifier, \lq{\textit{Material\_ID}}\rq. These nodal alloy compositions and their properties from $\tau_{\text{CGA}}$ are enumerated in Tab.~\ref{tab:tree_compositions} ordered from left to right based on Fig.~\ref{fig:gradient_tree}, and specifically identified by $Material\_ID$. As a result of the graph construction, this gradient is composed of single-phase BCC alloys with $M_{\text{creep}}$ and $\sigma_\text{y}$ entirely above the bottom quartile. Furthermore, as a result of the cost function, this gradient has a low worst-case cracking susceptibility $\text{Kou}' = 0.066$ for a non-terminal alloy in the $\tau_\text{CGA}$, with total Euclidean compositional edge length of 1.20 mole fraction [n/n]. Meanwhile, simply using Euclidean composition distance as the cost function results in a higher worst-case value of $\text{Kou}' = 0.079$, but a shorter total Euclidean compositional edge length of 0.57 mole fraction [n/n]. This indicates that the cost function works as intended to achieve the desired trade-off between length and worst-case cracking susceptibility.

\begin{table}
\renewcommand{\arraystretch}{1.3}
\centering
\caption{\textbf{Gradient Alloy Tree Nodal Compositions \& Relevant Properties}}
\label{tab:tree_compositions}
\begin{tabular}{|c|c|c|c|c|c|}
\hline
\textbf{$Material\_ID$} & \textbf{Composition} & \textbf{$\sigma_\text{y}$ {[}MPa{]}} & \textbf{$M_{\text{creep}}$ [m$^{-2}$ s]} & \textbf{$\frac{B}{S}$} & \textbf{$\text{Kou}'$} \\ \hline
2840\footnotemark[1]$^{,}$\footnotemark[2] & Cr$_{30}$Nb$_{5}$V$_{40}$W$_{25}$ & 786 & 429 & 2.18 & 0.077 \\ \hline
1195 & Cr$_{30}$V$_{45}$W$_{25}$ & 580 & 576 & 2.16 & 0.027 \\ \hline
1200 & Cr$_{30}$V$_{50}$W$_{20}$ & 433 & 247 & 2.21 & 0.024 \\ \hline
1305 & Cr$_{35}$V$_{45}$W$_{20}$ & 479 & 243 & 2.22 & 0.022 \\ \hline
1080 & Cr$_{25}$V$_{55}$W$_{20}$ & 386 & 254 & 2.12 & 0.026 \\ \hline
1396 & Cr$_{40}$V$_{40}$W$_{20}$ & 523 & 241 & 2.32 & 0.021 \\ \hline
944 & Cr$_{20}$V$_{60}$W$_{20}$ & 337 & 266 & 2.44 & 0.028 \\ \hline
1474\footnotemark[1] & Cr$_{45}$V$_{35}$W$_{20}$ & 564 & 242 & 1.94 & 0.020 \\ \hline
948 & Cr$_{20}$V$_{65}$W$_{15}$ & 225 & 121 & 2.52 & 0.022 \\ \hline
795 & Cr$_{15}$V$_{70}$W$_{15}$ & 187 & 131 & 2.65 & 0.024 \\ \hline
2335 & Cr$_{10}$Nb$_{5}$V$_{70}$W$_{15}$ & 282 & 159 & 2.83 & 0.043 \\ \hline
3665 & Cr$_{5}$Nb$_{10}$V$_{70}$W$_{15}$ & 337 & 171 & 3.03 & 0.043 \\ \hline
4958 & Cr$_{5}$Nb$_{15}$V$_{65}$W$_{15}$ & 422 & 171 & 3.07 & 0.042 \\ \hline
6063 & Cr$_{5}$Nb$_{20}$V$_{50}$W$_{15}$ & 488 & 177 & 3.10 & 0.041 \\ \hline
6183 & Cr$_{10}$Nb$_{20}$V$_{55}$W$_{15}$ & 565 & 159 & 2.92 & 0.047 \\ \hline
6179 & Cr$_{10}$Nb$_{20}$V$_{50}$W$_{20}$ & 681 & 327 & 2.83 & 0.060 \\ \hline
6284 & Cr$_{15}$Nb$_{20}$V$_{45}$W$_{20}$ & 770 & 297 & 2.68 & 0.066 \\ \hline
7191\footnotemark[1] & Cr$_{15}$Nb$_{25}$V$_{40}$W$_{20}$ & 830 & 314 & 2.71 & 0.067 \\ \hline
\end{tabular}
\footnotetext[1]{Terminal Alloy Node}
\footnotetext[2]{Root Node of Tree}
\renewcommand{\arraystretch}{1.0}
\end{table}

\subsubsection{Conformal Compositional Gradient Mapping \& Final Design}
Finally, with a gradient tree of alloy compositions identified, the multi-terminal gradient was ready for placement within the part geometry. In order to conformally place this gradient within a part geometry, the TreeMAP algorithm was used to maximize the use of the creep-resistant $x_{B}$ in the interior of the blade after the required gradient layers are placed. With this approach, a spatial part graph, $G_{P} = (V_{P}, E_{P})$, corresponding to a discretized 3D model of our part is necessary for efficient traversal. This graph has nodes, $V_{P}$, corresponding to each unique discrete geometric cell, and edges, $E_{P}$, between geometrically adjacent cells. A $Material\_ID$ property was applied to the nodes $V_{P}$ to indicate the geometric placement of materials based on their general node index in the ATLAS materials graph $G_{\text{AM}}$.

For the 3D model, a publicly available STL file of a gas turbine blade from GrabCAD was used \cite{grabcad_gas_turbine_blade}. CAD software (ANSYS\textsuperscript{\textregistered} SpaceClaim\textsuperscript{\textregistered} Direct Modeler\textsuperscript{\texttrademark}, version 2023) was then used to process the STL. To develop a discretized representation, the STL surface mesh was then converted into a voxelization with axial step sizes of $(\text{d}x, \text{d}y, \text{d}z) = (0.4, 0.4, 1.0)$ in millimeters using the \textit{pyvista.voxelize} function from PyVista \cite{sullivanPyVista3DPlotting2019} Python interface for the Visualization Toolkit (VTK) \cite{vtkBook}. The voxel step sizes were based on the reported hatch spacing and the compositional resolution in the build direction for multi-material AM with steels using LP-DED by Salas et al. \cite{salasEmbeddingHiddenInformation2022a}. Here, the $x$- and $y$-axis resolutions were based on the reported hatch spacing, and the $z$-resolution was based on the reported number of layers in the build direction to achieve a distinct compositional band. The resultant voxelization, stored as an unstructured PyVista grid, contained $2.49 \times 10^{6}$ cells. This was then converted into a spatial part graph $G_{P}$ for input to the TreeMAP algorithm. First, the graph $G_{P}$ was defined, with corresponding nodes, $V_{P}$, for each voxel cell and edges, $E_{P}$, between cells that share a vertex in the unstructured grid. Constructing $G_{P}$ for the gas turbine blade resulted in a graph with $2.49 \times 10^{6}$ nodes and $3.13 \times 10^{7}$ edges.

The terminal alloy for the surface of the turbine blade, $x_{A}$ was located within the part geometry by using PyVista to import a mesh file encompassing the blade, which translates to $z > -3$ in the STL mesh file coordinates. This mesh was used to select the voxels in the entire blade $z > -3$, which were then down-selected to only the exterior surface voxels. These surface voxels on the blade were labeled with $Material\_ID = 1474$ for $x_{A}$.

The terminal alloy for the base of the turbine blade, $x_{B}$, was located based on the location of $x_{A}$ and the number of edges (16) in the tree between $x_{A}$ \& $x_{B}$. All voxels more than 16 edges away from $x_{A}$ in the part graph $G_{P}$, and with $z < -3$ (to avoid encroaching on the blade core), were assigned $x_{B}$ with $Material\_ID = 7191$.

The placement of the terminal alloy $x_{C}$ could be determined similarly to $x_{B}$. However, in this work, we defined the so-called \lq{coalescent material}\rq~TreeMAP algorithm parameter $m_{c} = x_{C}$, with $Material\_ID = 2840$, allowing the algorithm to place the required conformal gradient layers and fill the remaining unlabeled voxels with $x_{C}$ as the last step to completing the gradient. Lastly, all unselected nodes were left unlabeled for TreeMAP to complete the conformal multi-terminal CGA mapping.

With $G_{P}$, $m_c$, and the gradient given as $\tau_{\text{CGA}}$, the TreeMAP algorithm was used to automatically map the multi-terminal CGA onto the discretized part geometry. Since one of the overarching goals was to maximize the usage/size of the high-temperature creep-resistant material, $x_{C}$ was used as the input for $m_{c}$. Once the necessary gradient layers are placed, TreeMAP fills any remaining unlabeled voxels by assigning $m_{c} = x_{C}$ as the material. In essence, this amounts to maximizing the usage of the high-performing alloy $x_{C}$ in the interior of the part, including the core, where its high-temperature creep and strength properties are particularly desirable. While using $m_{c} = x_{C}$ may result in $x_{C}$ being used elsewhere in the part, $x_{C}$ possesses notable mechanical properties compared to the other compositions in the gradient, so it is desirable to use it rather than distribute the gradient compositions with subpar material properties over a larger physical area. 

\begin{figure}
\centering
\includegraphics[width=\textwidth]{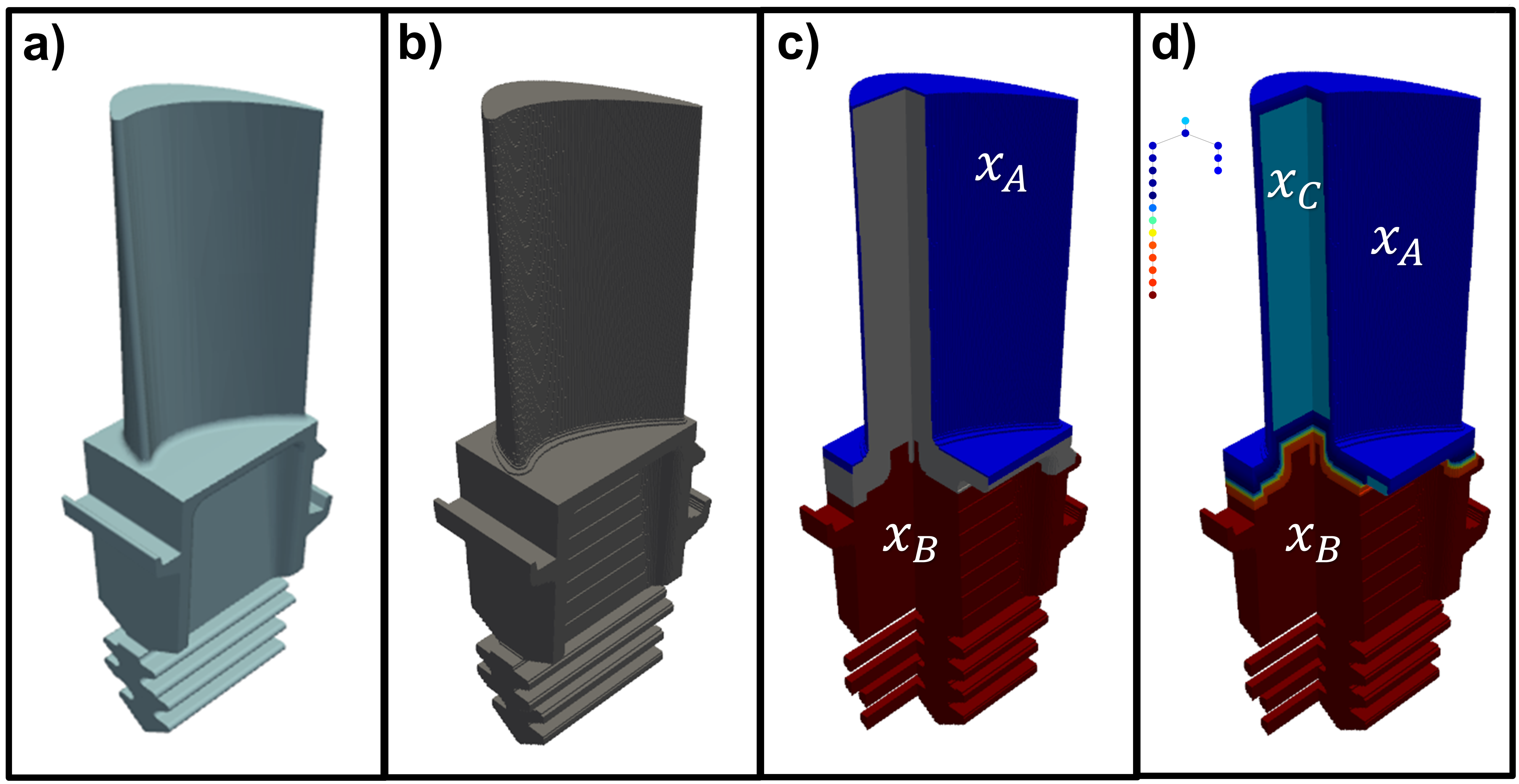}
\caption{\textbf{Illustration of the gas turbine blade throughout the design process. a)} the initial STL 3D surface mesh, \textbf{b)} the volumetrically-discretized, unlabeled model from PyVista, \textbf{c)} a cutaway of the discretized model, partially labeled with the spatial placement of terminal alloy compositions $x_{A}$ \& $x_{B}$, and \textbf{d)} a cutaway of the final discretized model with conformal CGA placement, colored based on the corresponding $Material\_ID$ from the gradient tree, $\tau_{\text{CGA}}$, of alloy compositions.}\label{fig:part_evolution}
\end{figure}

As a result, $G_{P}$ was labeled for a compositionally graded alloy between the three terminal compositions in a CPU time of 38.05 seconds using 32 Intel Xeon CPU cores at 2.20 GHz and 51.0 GB of system RAM. Mapping the resultant, labeled voxel nodes $V_P \in G_{P}$ back to the PyVista voxelization resulted in the final CGA structure. For the materials in the resultant gas turbine blade design, shown in a cutaway in Fig.~\ref{fig:part_evolution}d and in orthogonal cross-sections in Fig.~\ref{fig:part_design}a by $Material\_ID$, the terminal alloy compositions can be seen in their prescribed locations, as well as the conformal layers of alloy compositions from $\tau_{\text{CGA}}$. Figure ~\ref{fig:part_evolution}a-d illustrates the evolution of the part design from the initial STL 3D model representation to the final compositionally graded gas turbine blade structure. Note that the gradient from $x_A$ to $x_B$ largely bypasses $x_C$ due to their initial placement based on the required gradient steps; $x_C$ only appears where there is additional space after placing all other gradient compositions.

\subsection{Discussion}
\begin{figure}
\centering
\includegraphics[width=\textwidth]{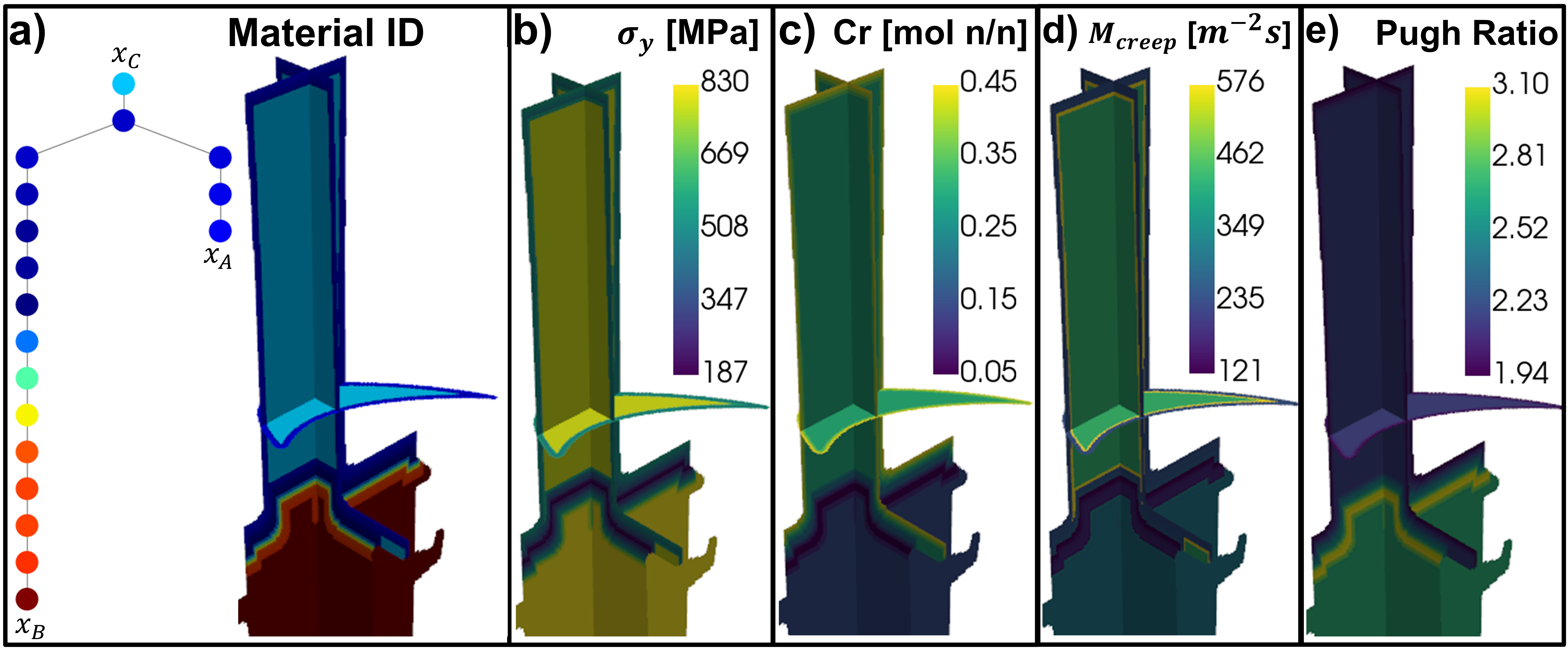}
\caption{\textbf{Orthogonal section views showing the spatial distribution of the alloys and properties for the voxels in the final turbine blade design. a)} $Material\_ID$, \textbf{b)} yield strength, \textbf{c)} chromium content, \textbf{d)} creep merit index, and \textbf{e)} Pugh ratio}\label{fig:part_design}
\end{figure}

Recalling the original goals for this design outlined in Fig.~\ref{fig:design_problem_definition}, Fig.~\ref{fig:part_design} shows the resulting design from the performance-driven approach taken in this work. First and foremost, Fig.~\ref{fig:part_design}a shows the unique materials by their $Material\_ID$ and indicates that the materials in the part follow the adjacency required by the Steiner tree to ensure single BCC phases, avoiding both deleterious phases and any other undesirable (non-BCC) phases. In Fig.~\ref{fig:part_design}b, it is evident that $\sigma_{\text{y}}$ does not fall below the 25th percentile value of 167 MPa. Here we also see that $x_{C}$ in the blade core and $x_{B}$ in the base both possess an extremely high yield strength, above the 95th percentile value of 771 MPa. For oxidation resistance, Fig.~\ref{fig:part_design}c highlights the placement of $x_{A}$ at the blade surface with 45 at. \% Cr, which is dramatically higher than the other compositions in the gradient. Next, Fig.~\ref{fig:part_design}d indicates the high creep merit index of $x_{C}$. While not as high as some other alloys in the gradient, this value is noteworthy due to the significant yield strength of this alloy. Lastly, Fig.~\ref{fig:part_design}e indicates a similar situation for $x_{B}$ concerning the Pugh ratio. Its magnitude is not the highest, but it is noteworthy when considering the exceptional yield strength of the alloy. Overall, the \textbf{resultant design has a combination of properties that would not be possible if limited to one alloy composition for conventional manufacturing}. These results indicate the utility of the presented workflow for performance-driven design of compositionally graded metal parts with enhanced performance compared to parts comprised of a single alloy, and extends the consideration of AM for simplifying assemblies even when they are comprised of parts with different alloy compositions.

In addition to observing the final design in light of the design intent, we can also observe some new considerations for CGAs from Fig.~\ref{fig:part_design} concerning the design of physical parts. It is evident from the cross-sections shown in Fig.~\ref{fig:part_design} that the number of material steps in a gradient translates to more physical space required, illustrated by the small gradient region between $x_{A}$ \& $x_{C}$ compared to $x_{B}$ \& $x_{C}$. This indicates the practical value of the simple cost function used for several early works on computational CGA design \cite{kirkComputationalDesignGradient2018, kirkComputationalDesignCompositionally2020a}, which is the Euclidean compositional distance. While barycentric coordinates are often used for alloy composition spaces, the compositions should be represented in terms of all Euclidean elemental components to enforce consistent lengths across the alloy composition space. This computation is functionally equivalent for Euclidean and barycentric coordinates if the simplicial vertices are pure elements, which is not always the case. Also, any balance element coordinate must be included to avoid skewing the compositional distances. The Euclidean compositional distance cost $c_d$ can be computed for all edges $e_{J,i} \in E_{J}$ as:

\begin{equation}
\label{eq:length_cost}
    c_d(e_{J,i}) = \|\mathbf{x_{j}} - \mathbf{x_{k}}\|_2
\end{equation}

\noindent where $c_d$ is the weight for an arbitrary graph edge $e_{J,i} = \left(j, k\right) \in E_J$, $x_{j}$ and $x_{k}$ are the elemental Euclidean compositions at the $Alloy$ nodes $v_{j}$ and $v_{k}$ respectively, and \(\|\cdot\|_2\) represents the Euclidean norm.  This cost function minimizes the total change in composition, translating to less unique material layers required to print a given CGA. Had the gradient between $x_{A}$ \& $x_{B}$ required many more gradient material steps, $x_{B}$ may have occupied an inconsequential amount of space on the interior of the blade, if there was any space at all. Since the discretization was based on the spatial, compositional resolution of a real metal AM machine \cite{salasEmbeddingHiddenInformation2022a}, this also points to the benefit of using AM technologies with finer printing resolution, as these (future) technologies could deposit compositional gradients with a given number of discrete steps over a smaller physical volume. 

Overall, the results show that the combined capabilities of recent methods can be used for the performance-driven design of terminal alloys, multi-terminal CGAs, and complex, conformally compositionally graded alloy structures with enhanced performance capabilities. Based on the materials data collected, these capabilities would be unattainable using any other single-phase BCC alloy composition in this alloy system. If current efforts \cite{sbir_2651311} to develop complementary capabilities in manufacturing technology are successful, these structures may soon become a physical reality.

\section{Methods}\label{sec:methods}

\subsection{Simplicial Grid Sampling}
The first step in assessing the alloy composition space for an ATLAS materials graph is acquiring a compositional sampling. For this work, a simplicial grid graph was efficiently generated for the Cr-Nb-V-W-Zr quinary system with a sample spacing of 0.05 mole fraction [n/n] using the NIM simPLEX (NIMPLEX) Nim library for efficient generation of compositional grid samplings and graphs \cite{krajewskiEfficientGenerationGrids2024b}. This sampling resulted in 10,626 unique alloy compositions and associated local grid edges.

\subsection{Thermodynamic Calculations} 

Thermodynamic calculations were performed at the sampled compositions using Thermo-Calc \cite{andersson2002thermo}, a CALPHAD-based software package, to explore the thermodynamic behavior of the alloy systems under study. Both equilibrium and Scheil-Gulliver solidification \cite{Scheil+1942+70+72, gulliver1913quantitative} calculations were employed to map out the alloys' feasible thermodynamic space and solidification pathways. The Scheil-Gulliver solidification model describes a non-equilibrium process where complete mixing occurs in the liquid phase, but diffusion in the solid phase is neglected. As a result, solidification proceeds progressively, with the composition of the remaining liquid changing continuously according to the partition coefficient. In this model, the solute is rejected into the liquid as solid forms, leading to segregation and compositional gradients within the material. This results in an inhomogeneous solid, where different regions solidify at different temperatures, broadening the solidification range.

\subsubsection{Equilibrium Calculations} 
Equilibrium phase simulations were conducted for each composition using Thermo-Calc \cite{andersson2002thermo} version 2023b with the TCHEA6 database \cite{TCHEA6}. We sampled temperatures from $1000^\circ C$ to $2750^\circ C$ at intervals of $250^\circ C$ for each composition, and only BCC phases were collected. Regarding the accuracy of the TCHEA6 database, in prior work \cite{kirk2022entropy}, the TCHEA4 database was benchmarked against experimental data and found to offer acceptable accuracy. The TCHEA6 database, being more comprehensive with additional thermodynamic assessments, is expected to provide improved predictive accuracy.

\subsubsection{Scheil-Gulliver Solidification Calculations} 

Using Thermo-Calc to predict the Scheil-Gulliver curve, we obtained the Scheil solidification range, the Kou hot-cracking criterion, and the phases present upon the end of rapid solidification. The solidification range is a critical consideration when designing alloys that, to a first approximation, are resistant to hot-cracking and solidification defects \cite{sheikh2024exploring}. The equilibrium solidification range is often narrower than the more realistic solidification range that occurs during rapid cooling, such as in Scheil-Gulliver solidification \cite{Scheil+1942+70+72, gulliver1913quantitative}. The latter more accurately reflects non-equilibrium conditions, where incomplete diffusion in the solid phase leads to a broader solidification range. Furthermore, quantities derived from the Scheil-Gulliver curve are often used as hot-cracking indicators, such as the Kou cracking susceptibility criterion \cite{kou2015simple}. This metric is defined as the magnitude of the temperature derivative with respect to the square root of the solid fraction near the end of solidification, $\mid dT/d\sqrt{f} \mid$. Kou's criterion has been used as a hot cracking indicator in various alloy design scenarios, including CGAs \cite{kou2015simple, yangDesignMethodologyFunctionally2023a}. Finally, the Scheil-Gulliver phase fractions at the end of solidification were used as an estimate of what phases may form at the end of a rapid solidification processing pathway.

\subsection{Material Property Modeling} 
Various models were employed to ensure that the designed part met requirements for properties, including high-temperature yield strength, ductility, and resistance to creep. In this work, higher ductility was used as an indicator of increased fatigue life performance. Some of the most important property models are described below.

\subsubsection{Pugh Ratio}
To encode ductility requirements into the CGA design problem, Pugh ratio was used \cite{pughXCIIRelationsElastic1954}. This is a commonly employed design metric for ductility where higher values typically provide a useful indicator for ductile behavior in concentrated multi-component alloys \cite{singh2023ductility}. A Pugh ratio of $B/S{=}1.75$ is often defined as the transition from brittle to ductile behavior, where $B$ and $S$ are the bulk and shear moduli, respectively \cite{liu2020AlCrTiFeNi, liu2020CrFeCoNiMox, zhang2018elastic}. In the present work, these elastic constants were calculated for each composition in the design space via a Rule-of-Mixtures average of the pure elements' properties.

\subsubsection{Yield Strength Model}
A strengthening model for refractory BCC high entropy alloys originally proposed by Maresca and Curtin was used to predict high-temperature yield strength for this framework \cite{maresca2020MechanisticOriginOfStrength}. This model was identified as an appropriate estimate of strength because the present work uses a single-phase BCC alloy design space where alloys with multiple principal elements make up a significant portion of the feasible compositions. Alloy shear strength is approximated over a wide range of temperatures using
\begin{equation}
\tau_\text{y}(T, \dot{\varepsilon})=\tau_{\text{y} 0} \exp \left[-\frac{1}{0.55}\left(\frac{k T}{\Delta E_b} \ln \frac{\dot{\varepsilon}_0}{\dot{\varepsilon}}\right)^{0.91}\right],
\label{eq:Maresca-Curtin_Approx_a}
\end{equation}
where $k$ is the Boltzmann constant, $T$ is the absolute temperature at which yield stress is being predicted, $\tau_{\text{y}0}$ is the zero-temperature shear stress, and $\Delta E_b$ is the energy barrier for the motion of each dislocation segment. $\dot{\epsilon}$ and $\dot{\epsilon}_0$ are the applied strain rate, set at $\dot{\epsilon} = 10^{-3} s^{-1}$, and a reference strain rate, estimated to be $\dot{\epsilon}_0 = 10^4 s^{-1}$. This equation creates what has been shown to be a lower bound estimate \cite{baruffi2022screw, vela2023data} for tensile yield strength with
\begin{equation}
\sigma_\text{y}(T, \dot{\varepsilon})=M\tau_\text{y}(T, \dot{\varepsilon})
\label{eq:Maresca-Curtin_Approx_b}
\end{equation}
where $M=3$ was used as the Taylor factor of a BCC metal. Further detail on how to calculate the inputs to the equations can be found in \cite{acemi2024multi}. This includes a previously published description of the framework employed in this work. It also explains how Rule-of-Mixture averages of relevant elastic constants and volumes of the pure elements are used to estimate Maresca-Curtin yield strength.

\subsubsection{Creep Model}
The creep merit index $M_{\text{creep}}$ originally proposed by Reed et al. \cite{REED20095898} was used as a design indicator for creep resistance in the alloy space,
\begin{equation}
    M_{\text{creep}} = \sum_{i}{\frac{x_i}{\widetilde{D_i}}},
\end{equation}
where $x_i$ is the molar fraction of each component and $\widetilde{D_i}$ is the interdiffusion coefficient for that component. This is meant to capture the resistance to diffusion-based creep mechanisms such as Nabarro-Herring and Coble creep, where larger creep merit indices correlate with greater resistance to creep deformation. Diffusion coefficients were queried at 2273~K from Thermo-Calc's MOBHEA3 database with the reference element selected as the largest atomic radius in each composition.

\subsection{Deep Learning Property Regression Model} 
\label{sec:regressor_model}
\begin{figure*}[!ht]
            \centering
            \scriptsize
            \begin{overpic}[width=0.32\linewidth]{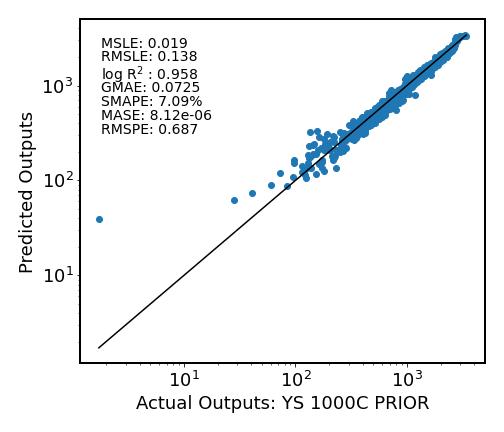}
                \put(60,15){\includegraphics[width=0.11\linewidth]{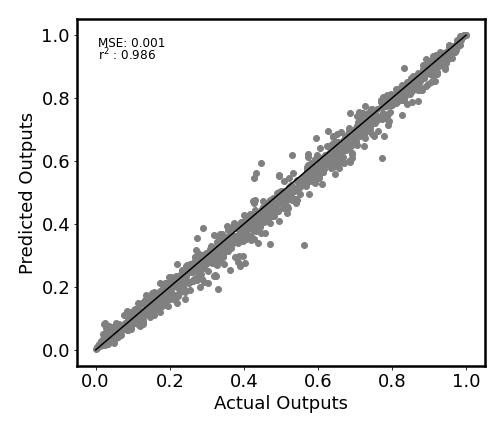}}
                \put(1,80){a)}
            \end{overpic}
            \begin{overpic}[width=0.32\linewidth]{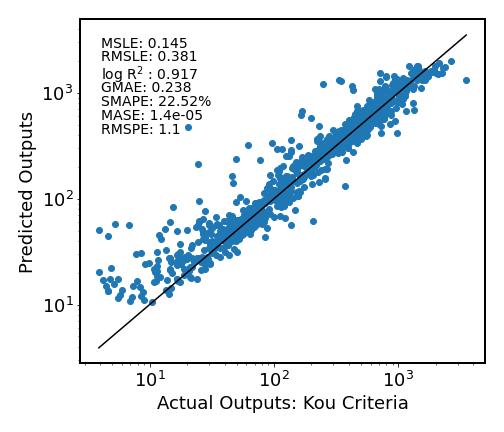}
                \put(60,15){\includegraphics[width=0.11\linewidth]{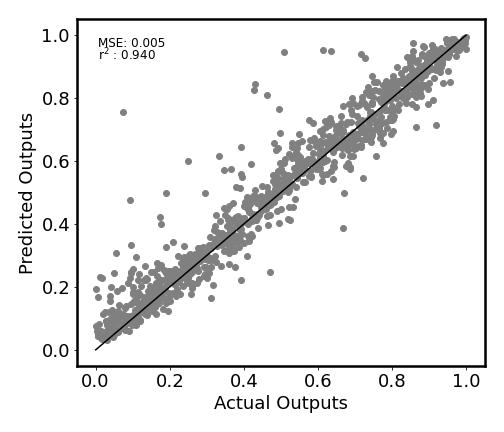}}
                \put(1,80){b)}
            \end{overpic}  
            \begin{overpic}[width=0.32\linewidth]{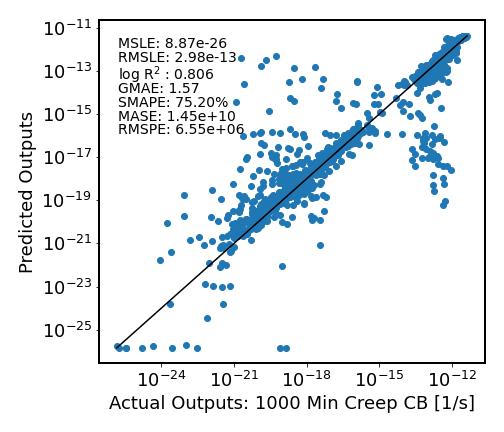}
                \put(60,15){\includegraphics[width=0.11\linewidth]{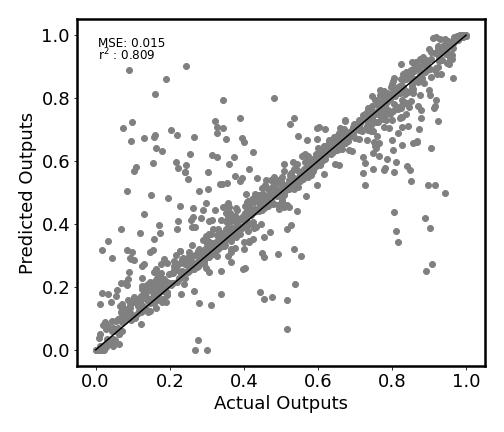}}
                \put(1,80){c)}
            \end{overpic}
            \\
            \begin{overpic}[width=0.32\linewidth]{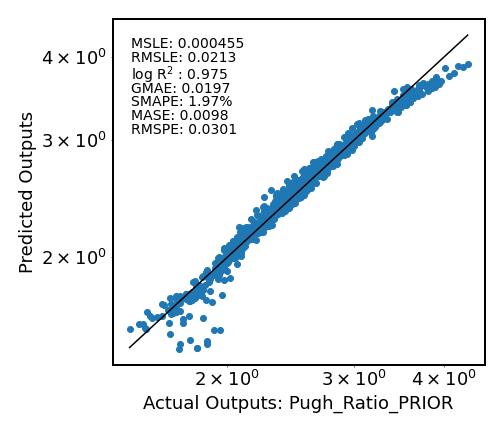}
                \put(60,15){\includegraphics[width=0.11\linewidth]{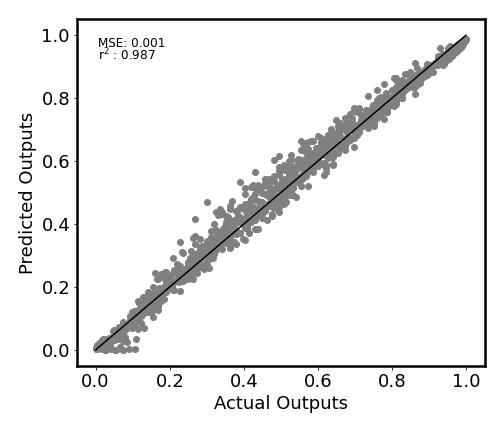}}
                \put(1,80){d)}
            \end{overpic}  
            \begin{overpic}[width=0.32\linewidth]{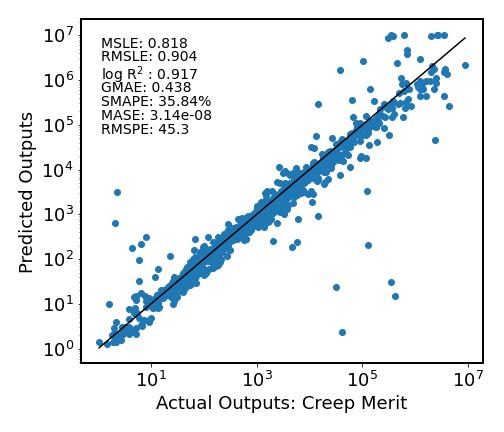}
                \put(60,15){\includegraphics[width=0.11\linewidth]{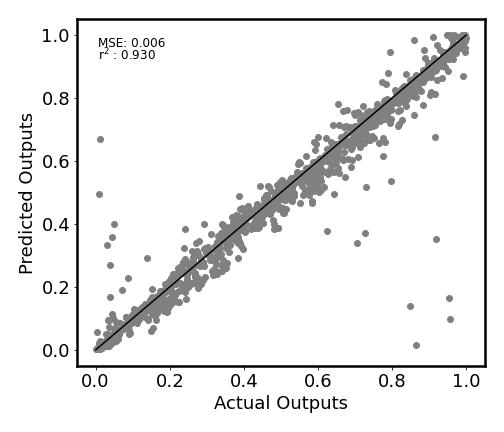}}
                \put(1,80){e)}
            \end{overpic}  
            \begin{overpic}[width=0.325\linewidth]{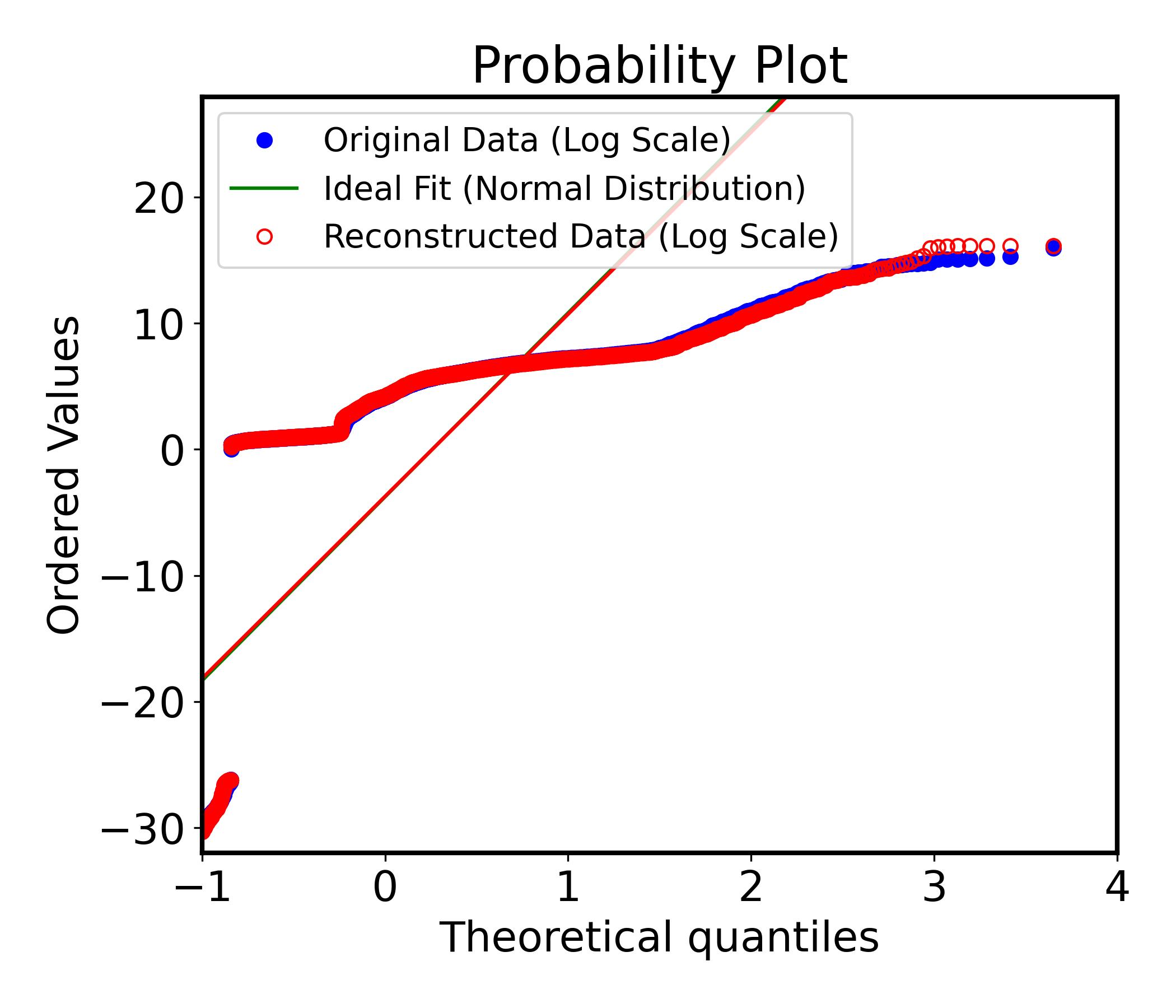}
                \put(33,14){\includegraphics[width=0.19\linewidth]{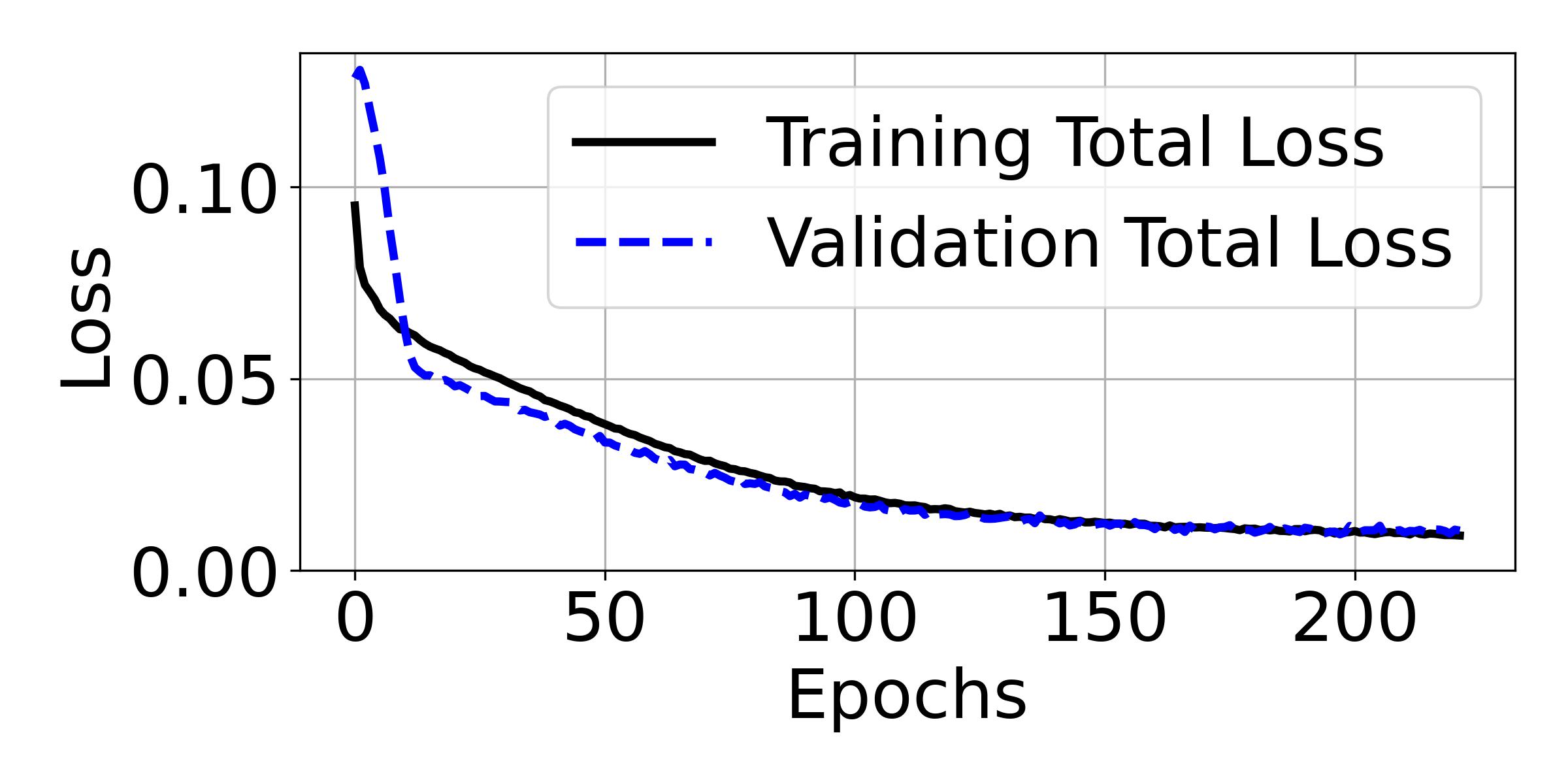}}
                \put(1,80){f)}
            \end{overpic}  
        \caption{\textbf{Prediction performance and probability analysis of the deep regressor for various material properties.} Parity plots for \textbf{(a)} Yield Strength (YS) at 1000$^\circ$C, \textbf{(b)} Kou Criteria, \textbf{(c)} Coble Creep at 1000$^\circ$C, \textbf{(d)} Pugh Ratio, and \textbf{(e)} Creep Merit showing high accuracy between predicted and actual points with strong agreement across different magnitudes. The small plots in each figure show the model performance for scaled data. \textbf{(f)} Quantile-Quantile plot comparing true and predicted values with an ideal fit for a vector of all features in log scale showing the model’s effectiveness in reconstructing the true data. The inset shows the history of training and validation loss throughout the training process.}\label{fig:fast_screening_properties}
\end{figure*}

We also employed a deep-learning property regression model \cite{attari2024decoding} to predict property values where needed. The model was trained on a dataset composed of compositional features as input variables (i.e., Cr, Nb, V, W, Zr) and target properties (e.g., yield strength at various temperatures, Pugh ratio, Coble min. creep rate \cite{cobleCreep1963}, creep merit index, and Kou criterion) as output variables. Specifically, an encoder component of the model transforms the input compositional data into a latent space, and a decoder component of the model, which mirrors the encoder's structure, maps this latent space to property features. This model is outlined in further detail in the Appendix~\ref{secA1}. This model captured non-linear relationships between composition and a variety of properties, as indicated by the prediction performance and probability analysis shown in Fig.~\ref{fig:fast_screening_properties}. In this work, we specifically used the model for predicting missing/erroneous values of Yield Strength at $1000^\circ$C and the Kou cracking susceptibility criterion at nodal $Alloy$ compositions. We also used it to predict the Kou cracking susceptibility criterion across the $JOINS\_TO$ edges for a property-based cost function discussed later.

\subsection{ATLAS Materials Graph Initialization} 
\subsubsection{Initial Data Structuring}

A graph developed using ATLAS captures the topology of the alloy composition space in a so-called \lq{ATLAS materials graph}\rq, $G_{\text{AM}} = (V_{A}, E_J)$, where unique alloy compositions are represented as $Alloy$ nodes, $V_A$, connected with $JOINS\_TO$ edges, $E_J$. A $JOINS\_TO$ edge is assigned to local composition pairs that could be joined in a compositional gradient. For example, in this work, we assumed that in a simplicial grid of alloy compositions, those separated by one grid step could be joined to one another, similar to other recent work on building graph representations of the alloy composition space \cite{krajewskiEfficientGenerationGrids2024b}.

The nodal alloy compositions generated with the NIMPLEX library \cite{krajewskiEfficientGenerationGrids2024b} were used for all phase and property predictions. Since the 10,626 resultant compositions can be dealt with in local memory, this sampling was used to initialize an ATLAS materials graph $G_{\text{AM}} = (V_A, E_J)$.

From the gathered phase data, we identified the single-phase BCC $Alloy$ compositions based on the predictions from the Scheil-Gulliver solidification and equilibrium simulations. For equilibrium simulations, we considered the phase predictions from the $T_{s}$ of the alloy down to $1000^\circ$C, assuming the kinetics below $1000^\circ$C are frozen out. We required $>99$ at. \% BCC in the predicted phases from these equilibrium and Scheil-Gulliver simulations. Before filtering these single-phase BCC compositions using the property data, two compositions with missing values for $\sigma_{\text{y}}$ were imputed using the regression model. We then filtered the single phase BCC $Alloy$ nodes further based on properties, limiting our selection to that met the global property constraints $M_{\text{creep}} > 43.4$ and $\sigma_{\text{y}} > 167$. By retaining these feasible $Alloy$ nodes and their $JOINS\_TO$ edges, we had a constrained ATLAS materials graph representation of the feasible alloy composition space for this design problem.

When the $Alloy$ nodes are filtered based on manufacturing constraints for a given processing method, the remaining $Alloy$ nodes and their associated $JOINS\_TO$ edges discretely represent a feasible alloy chemistry design space for that processing regime. Additionally, the remaining $JOINS\_TO$ edges specifically relate to composition pairs that can be joined with the given processing regime. The ATLAS also defines a higher level \lq{ATLAS graph}\rq , $G_A$, that uses several categorical node types that can efficiently filter $Alloy$ nodes using graph traversals in large datasets, namely $Element$, $Phase$, and $ConstrainedSubgraph$, in an overarching graph database. However, we note that the dataset used in this work was not large enough to benefit from implementing these aspects of the schema.

\subsubsection{Cracking Susceptibility Cost Function}\label{sec:cost_function}
Recent work on the manufacturability of CGAs indicates the benefit of using cracking susceptibility indicators for designing crack-free gradients \cite{yangDesignMethodologyFunctionally2023a}. However, as indicators, they suggest susceptibility and do not directly predict occurrence. Thus, in this work, we developed a cost function for minimizing the worst-case magnitude of the normalized Kou cracking criterion \cite{kou2015simple} $\text{Kou}'$, with a secondary consideration for minimizing the path length. This cost function was defined as: 
\begin{equation}
\label{eq:Kou_cost}
    c(e) = \int_{x_a}^{x_b} \left( \text{Kou}'(x) \right)^{P} dx
\end{equation}
where $\text{Kou}'(x)$ is the normalized criterion as a function of alloy composition $x$, and $P$ is a user-defined power factor parameter. Here, $P$ scales the contribution of $\text{Kou}'(x)$ to the overall area under the curve proportional to the change in composition $x$. This can be used to ensure the planning process prioritizes compositional routes with nominally lower magnitudes of $\text{Kou}'$ regardless of compositional length, since marginal increases are penalized exponentially. Thus, as $P$ increased in magnitude, this cost function effectively minimizes the worst-case magnitude of $\text{Kou}'$ to improve the manufacturability of the gradient compositions. By integrating over composition, the compositional distance is still considered so that for two paths with similar overall magnitude for $\text{Kou}'$, the gradient with a smaller change in the composition will have a lower cost. This work used $P = 3$ to minimize the worst-case value of $\text{Kou}'$. We found that $P > 3$ began to over-prioritize granular improvement in $\text{Kou}'$ for a drastically longer compositional gradient. The cost for each edge was computed by 1) interpolating the alloy compositions along each remaining $JOINS\_TO$ edge with a step size of 0.1 mole \%, 2) predicting $\text{Kou}'$ using the deep learning regression model with the interpolated compositions as input, and 3) computing the edge cost from Eq.~\ref{eq:Kou_cost} using trapezoidal integration for the predicted \& scaled $\text{Kou}'$ values over their corresponding interpolated compositions.

\subsection{5D Composition Space Topological Partitioning \& Analysis}

Evaluating the properties of the remaining feasible $Alloy$ compositions indicated the possible material property trade-offs and the global distribution of properties across the alloy composition space. Topological partitioning of the composition space by identifying the maximally connected subgraphs \cite{blackMaximallyConnectedComponent2020} of the $G_{\text{AM}}$ also enabled evaluation of distinct constrained subgraphs for continuous compositional gradation between terminal alloys and their specific property trade-offs.

\begin{figure}[!ht]
\centering
\includegraphics[width=0.5\textwidth]{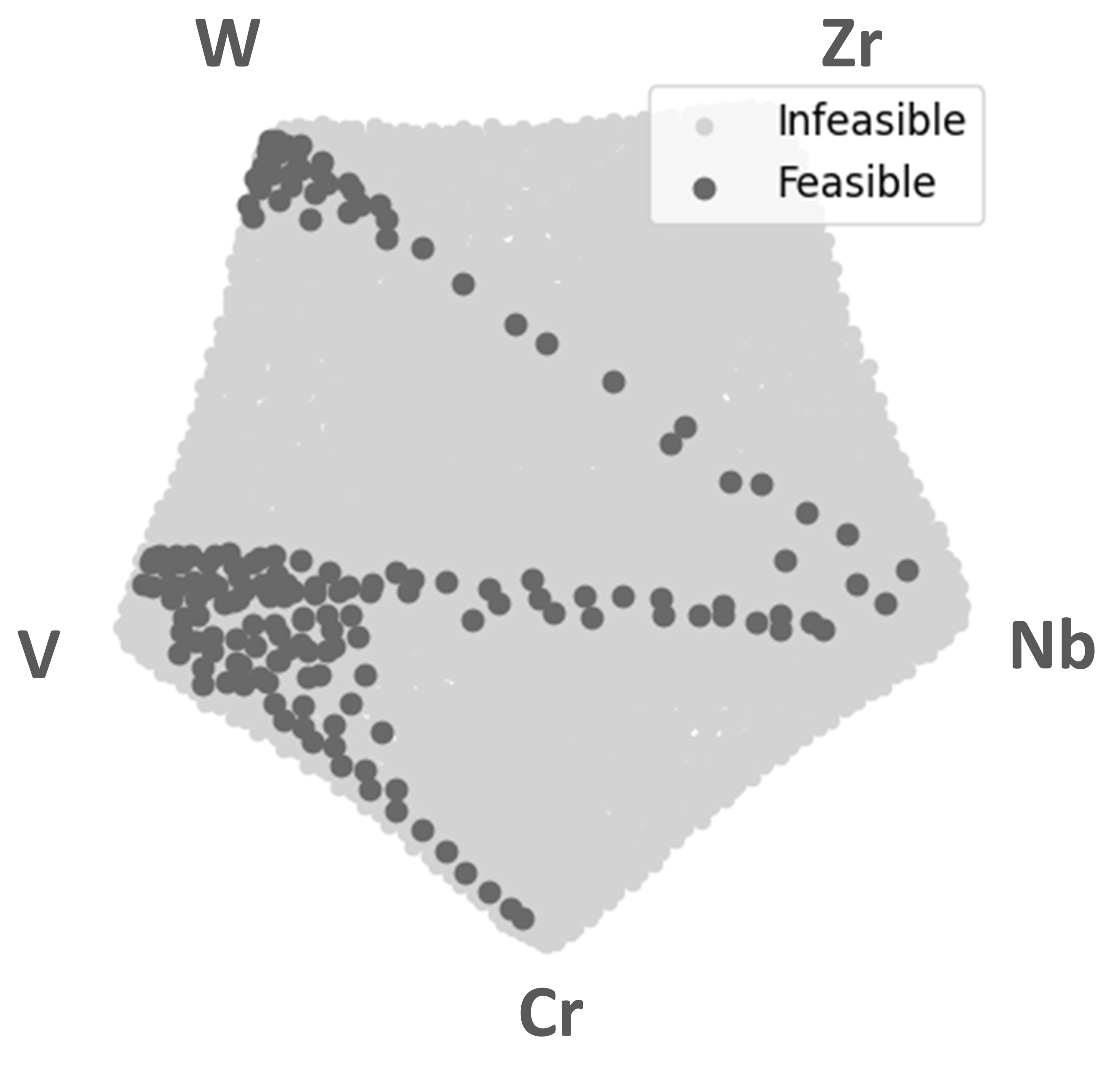}
\caption{\textbf{Uniform Manifold Approximation and Projection (UMAP) plot showing an embedded 2-dimensional representation of the infeasible alloy compositions (light grey) and feasible alloy compositions (dark gray) in the 5-component CrNbVWZr composition space.} In this embedding, similar alloy compositions are qualitatively displayed closer together. Alloy compositions with lower configurational entropy are displayed closer to the exterior boundary of the hypocycloid shape made by the embedding, while alloys with higher configurational entropy are displayed closer to the center. Pure elements are projected at the vertices, which are labeled with their corresponding element.}\label{fig:satisfactory-compositions-umap}
\end{figure}

\subsubsection{Feasible Composition Space}

\begin{figure}[!ht]
\centering
\includegraphics[width=0.99\textwidth]{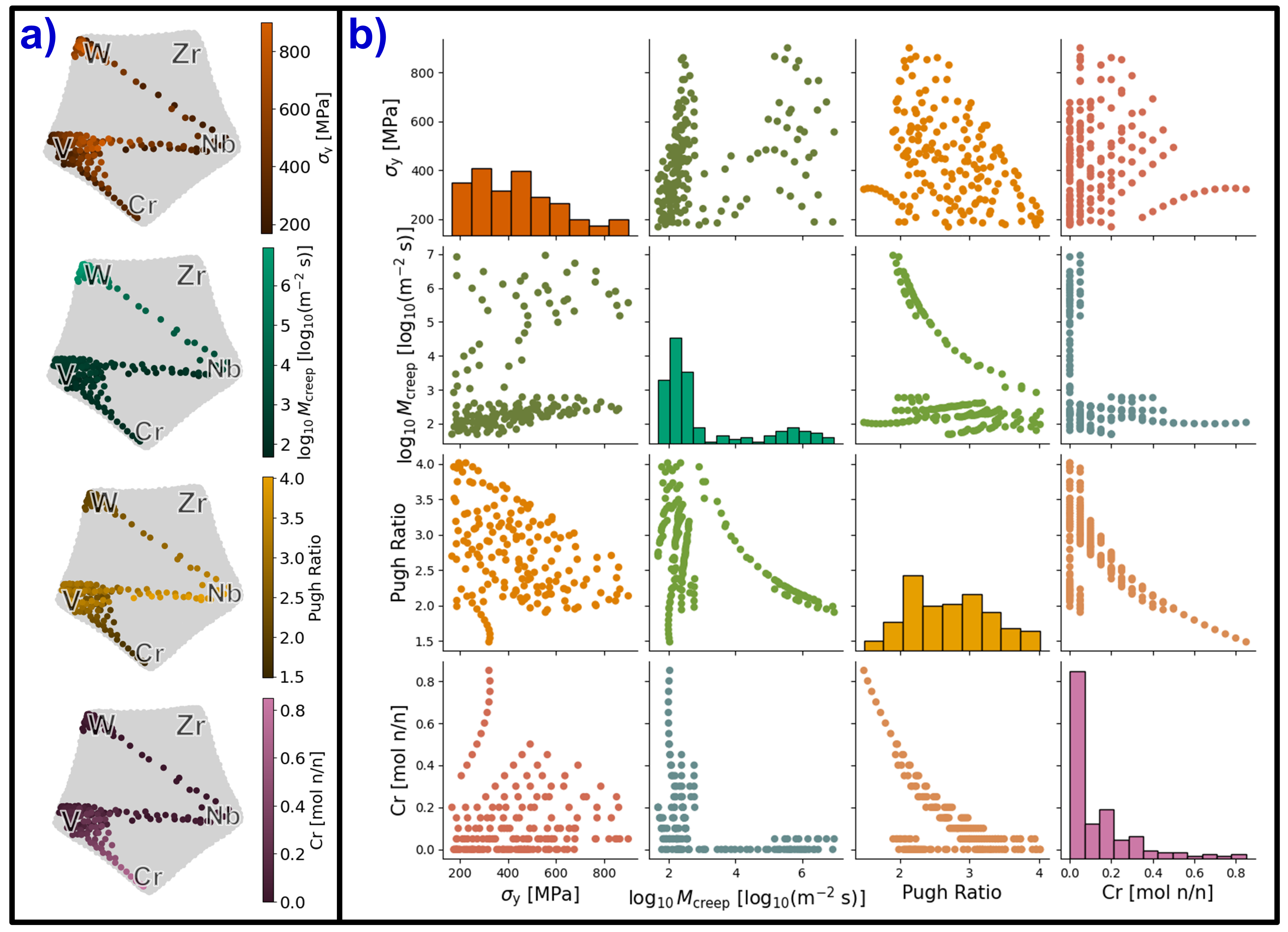}
\caption{\textbf{Relevant material properties for all feasible $Alloy$ nodal compositions in $G_{\text{AM}}$. a)} Uniform Manifold Approximation and Projection (UMAP) plots with color mapping illustrating the qualitative variation of the properties of interest with respect to alloy composition, and \textbf{b)} Scatterplot matrix of the properties of interest for this work, illustrating the pairwise correlation between properties (off-diagonal) and the distribution of each property (diagonal).}\label{fig:design_space_analysis}
\end{figure}

The materials graph representation can be used in conjunction with dimensionality-reduction methods like Uniform Manifold Approximation and Projection (UMAP) \cite{mcinnes2020umapuniformmanifoldapproximation} to visualize the feasible nodal $Alloy$ compositions from a high-dimensional chemistry space in 2D \cite{velaVisualizingHighEntropy2024}. Analyzing the feasible nodal $Alloy$ compositions $V_A$ for the feasible composition space in $G_{\text{AM}}$ revealed information about the chemistry of these compositions and their properties for materials design. Analyzing the filtered data indicated that the feasible alloy composition space contains compositions in 4 binary systems: CrW, NbV, NbW, and VW; 4 ternary systems: CrNbV, CrNbW, CrVW, and NbVW; and 2 quaternary systems: CrNbVW and NbVWZr. From this information, it was evident that Zr was only present in some feasible quaternary compositions and that no quinary compositions were feasible. 

With this data, UMAPs with different color mappings, shown in Fig.~\ref{fig:design_space_analysis}a, were used to qualitatively indicate the variation of the properties of interest with respect to alloy chemistry. From these plots, several trends were apparent. First, $\sigma_\text{y}$ appeared to decrease for compositions with higher V and Nb content, and increase for some W-rich compositions and compositions with relatively high configurational entropy near the V vertex. The creep merit index, ${M_{\text{creep}}}$, is shown on a log scale. $\log_{10}M_{\mathrm{creep}}$ appeared to increase as W content increases but does not show a significant trend with respect to the other elements. The Pugh ratio appears to be elevated for compositions containing higher amounts of Nb and V and lower for W and Cr. This trend is in agreement with the experimentally observed fact that Nb- and V-based alloys tend to be less brittle than W- and Cr-based ones. Lastly, the color map of Cr content indicates the presence of alloy compositions with moderate to high Cr content in the satisfactory alloy design space.

Meanwhile, the scatter plot matrix, shown in Fig.~\ref{fig:design_space_analysis}b, shows the distribution for each property in the histograms on the diagonal, as well as the pair-wise correlation between these properties in the scatter plots off the diagonal. From these plots, we observed left-skewed distributions for $\sigma_\text{y}$, $\log_{10}M_{\mathrm{creep}}$, and Pugh ratio. The Cr content is strongly left-skewed due to the presence of compositions with minimal to no Cr content. These plots make the property trade-offs between the available compositions more apparent. For example, in maximizing Cr content, the highest Cr content possible in the satisfactory alloy design space will be 0.85 mole fraction. 

\subsubsection{Constrained Subgraphs}

\begin{figure}
\centering
\includegraphics[width=0.5\textwidth]{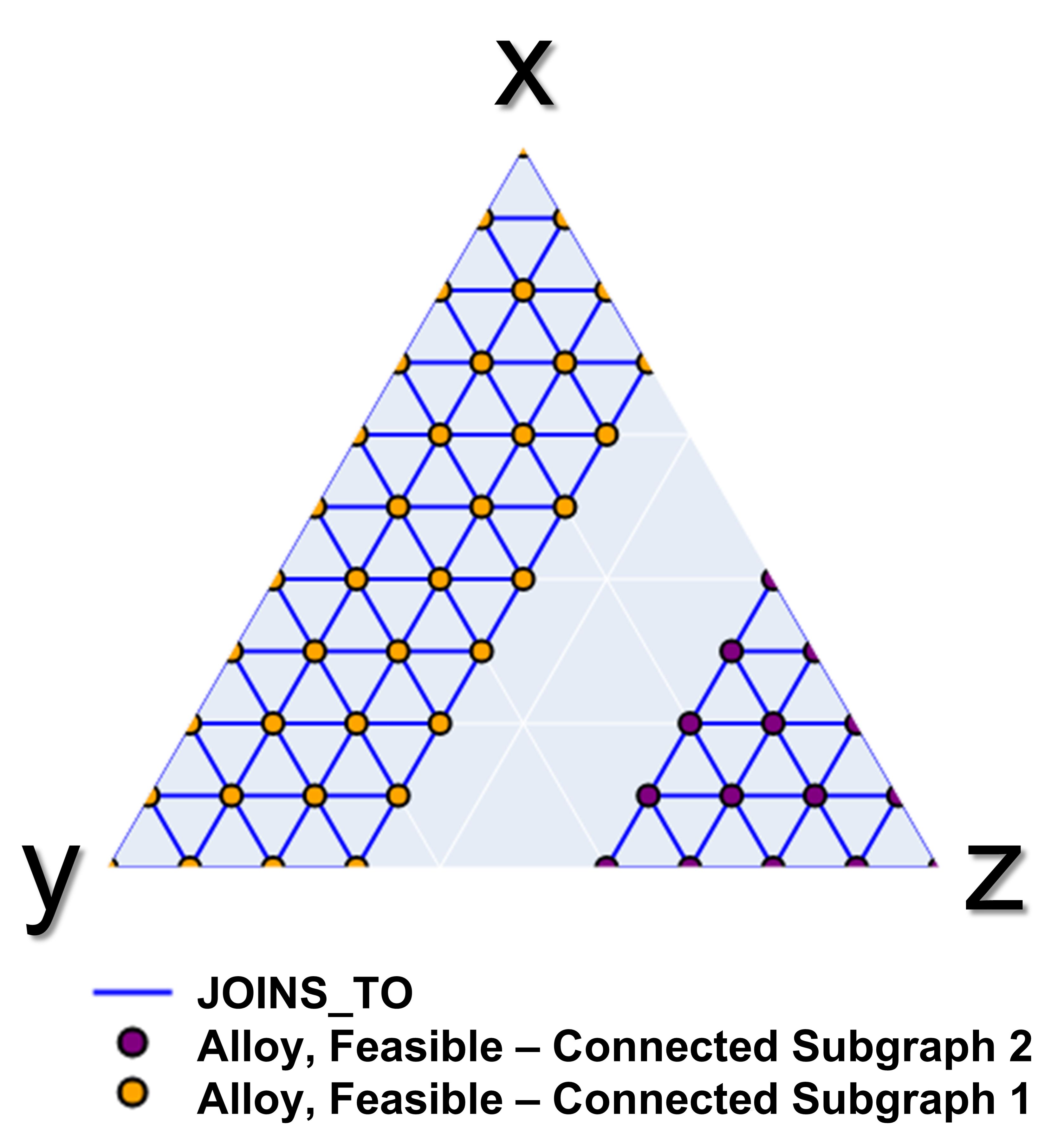}
\caption{\textbf{Illustration of graph reachability with constraint-isolated regions demonstrated in a synthetic ATLAS materials graph for a ternary alloy system comprised of elements \textbf{x}, \textbf{y}, and \textbf{z}.} For this graph, the $Alloy$ nodes have been filtered for feasibility, resulting in two maximally connected subgraphs indicated by purple and orange markers at the $Alloy$ nodes. Any unique pair of $Alloy$ nodes within a given subgraph can be connected by traversing the $JOINS\_TO$ edges for that subgraph. However, no pair of $Alloy$ nodes in different connected subgraphs can be connected by traversing $JOINS\_TO$ edges.}\label{fig:alloy_reachability}
\end{figure}

While this global analysis is useful, the topology of the feasible ATLAS materials graph must be considered, as there is no guarantee that a feasible gradient exists for any given pair of $Alloy$ nodes. To this end, we can interpret the topology of the feasible design space by partitioning the feasibility-constrained materials graph into maximally connected subgraphs \cite{blackMaximallyConnectedComponent2020}. In ATLAS, these are referred to as \lq{constrained subgraphs}\rq~since they are dependent on the specific constraints for a given design problem. Since these subgraphs are fundamentally disjoint, this partitioning guarantees the compatibility among pairs of nodal alloy compositions within a given constrained subgraph and the incompatibility of alloy nodes across different constrained subgraphs. Note that alloy compatibility can be considered reachable in the context of a graph; that is, for a given set of nodes, if some manufacturable gradient path over the edges through the graph exists between each unique pair, then these nodal alloy compositions are compatible. This concept is illustrated for a synthetic ATLAS materials graph in Fig.~\ref{fig:alloy_reachability}.

For the filtered ATLAS materials graph in this work, the constrained subgraphs were identified using the breadth-first-search (BFS) approach implemented in NetworkX for identifying maximally connected subgraphs \cite{hagbergExploringNetworkStructure2008}. Since every $Alloy$ node within a given subgraph is reachable from any other $Alloy$ within that subgraph using continuous compositional gradation along the $JOINS\_TO$ edges, these subgraphs can be used for the performance-driven selection of terminal alloys. Thus, the material property trade-offs available within each subgraph can be considered in isolation. This enables the selection of terminal alloy compositions for which a gradient of satisfactory compositions exists prior to any effort to design a specific, optimized gradient.

The four constrained subgraphs with the most $Alloy$ nodes contained 80, 40, 37, and 5 nodes, respectively. Only the two largest subgraphs contained compositions with notable amounts of Cr. We did not consider the third-largest subgraph due to our requirement for an oxidation-resistant (high-Cr) alloy, nor did we consider the fourth-largest subgraph or those smaller due to their insignificant size. 

As a result, the primary focus was directed towards the largest two subgraphs with 80 and 40 $Alloy$ nodes, which we will refer to as $G_{\alpha} = (V_{\alpha}, E_{\alpha})$ and $G_{\beta} = (V_{\beta}, E_{\beta})$ respectively, where $V_{\alpha}, V_{\beta} \subset V_{A}, ~\text{and}~ E_{\alpha}, E_{\beta} \subset E_{J}$. For these subgraphs, we performed the same analyses as the full satisfactory composition space. Observing the nodal alloy compositions for $G_{\alpha}$, we found that it contains compositions in three ternary systems: CrNbV, CrVW and NbVW; and one quaternary system: CrNbVW. Meanwhile, for $G_{\beta}$, we found that it contains compositions in two ternary systems: CrVW and NbVW; and one quaternary system: CrNbVW. Thus, Zr was not present in any amount in either subgraph.

\begin{figure}
\centering
\includegraphics[width=0.99\textwidth]{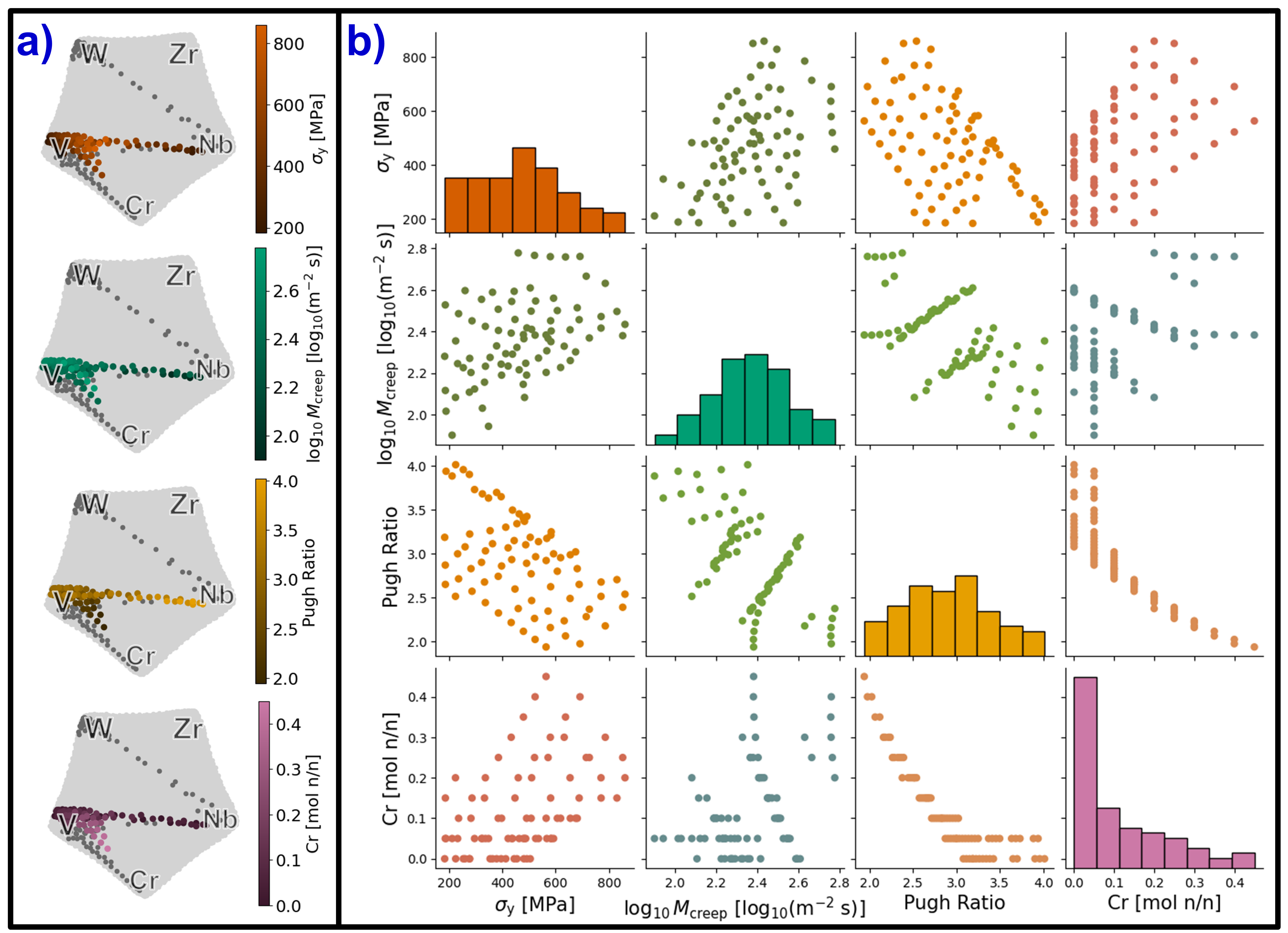}
\caption{\textbf{Relevant material properties for the $Alloy$ nodal compositions in the largest constrained subgraph $G_{\alpha}$. a)} color maps showing the qualitative variation of the properties of interest with respect to alloy composition using Uniform Manifold Approximation and Projection (UMAP) embedded coordinates, and \textbf{b)} Scatterplot matrix of the properties of interest for this work showing the pairwise correlation between properties (off-diagonal) and the distribution of each property (diagonal) for the reachable nodal compositions.}\label{fig:subgraph_1_space_analysis}
\end{figure}

The same UMAP embedding with different color maps, shown in Fig.~\ref{fig:subgraph_1_space_analysis}a, qualitatively indicated the variation of the properties of interest with alloy chemistry for the nodal alloy compositions in $G_{\alpha} \subset G_{\text{AM}}$. From these color maps, we saw that $\sigma_\text{y}$ appears to decrease for compositions with higher V or Cr content and less configurational entropy. Meanwhile, the $\log_{10}M_{\mathrm{creep}}$ increases for some compositions towards the V vertex. The Pugh ratio decreases for compositions containing higher amounts of Cr, and increases for higher Nb-content alloys. Lastly, the color map of Cr content indicates the presence of alloy compositions with moderate Cr content in this subset of the satisfactory design space, up to 0.45 mole fraction.

The scatter plot matrix for the nodal alloy compositions in $G_{\alpha}$, is shown in Fig.~\ref{fig:subgraph_1_space_analysis}b. From the plots in this matrix, we observed an approximately normal distribution for $\sigma_\text{y}$, with a range from approximately 150 to 875 MPa. In this subgraph, $\sigma_\text{y}$, $\log_{10}M_{\mathrm{creep}}$, and Pugh ratio are approximately normal in distribution. However, the Cr content is strongly left-skewed due to the presence of a compositions in the NbVW ternary with no Cr content. By partitioning the graph into constrained subgraphs, we now had \emph{the guarantee when evaluating these properties that all nodal alloy compositions were reachable from all other nodal alloy compositions} by traversing satisfactory alloy compositions. Since we leveraged the graph structure to achieve this reachability guarantee, we then observed the attainable property trade-offs for selecting terminal alloys for the compositionally graded part. From the off-diagonal scatter plots in Fig.~\ref{fig:subgraph_1_space_analysis}b, it was evident that we could select a composition with a moderate Cr content of 0.45 mole fraction, which is over twice the Cr content of many Ni-based superalloys. As a bonus, this can be achieved while maintaining a yield strength value above the average of 439 MPa among satisfactory compositions. Furthermore, some of the compositions with the highest value of the creep merit index in this subgraph also have well above-average values for $\sigma_\text{y}$. Lastly, we also observed that we can achieve a Pugh ratio well above the lower magnitude threshold of 1.75 typically used to indicate ductility \cite{pughXCIIRelationsElastic1954}, while achieving a yield strength above the 95th percentile threshold of 771 MPa or more.

\begin{figure}
\centering
\includegraphics[width=0.99\textwidth]{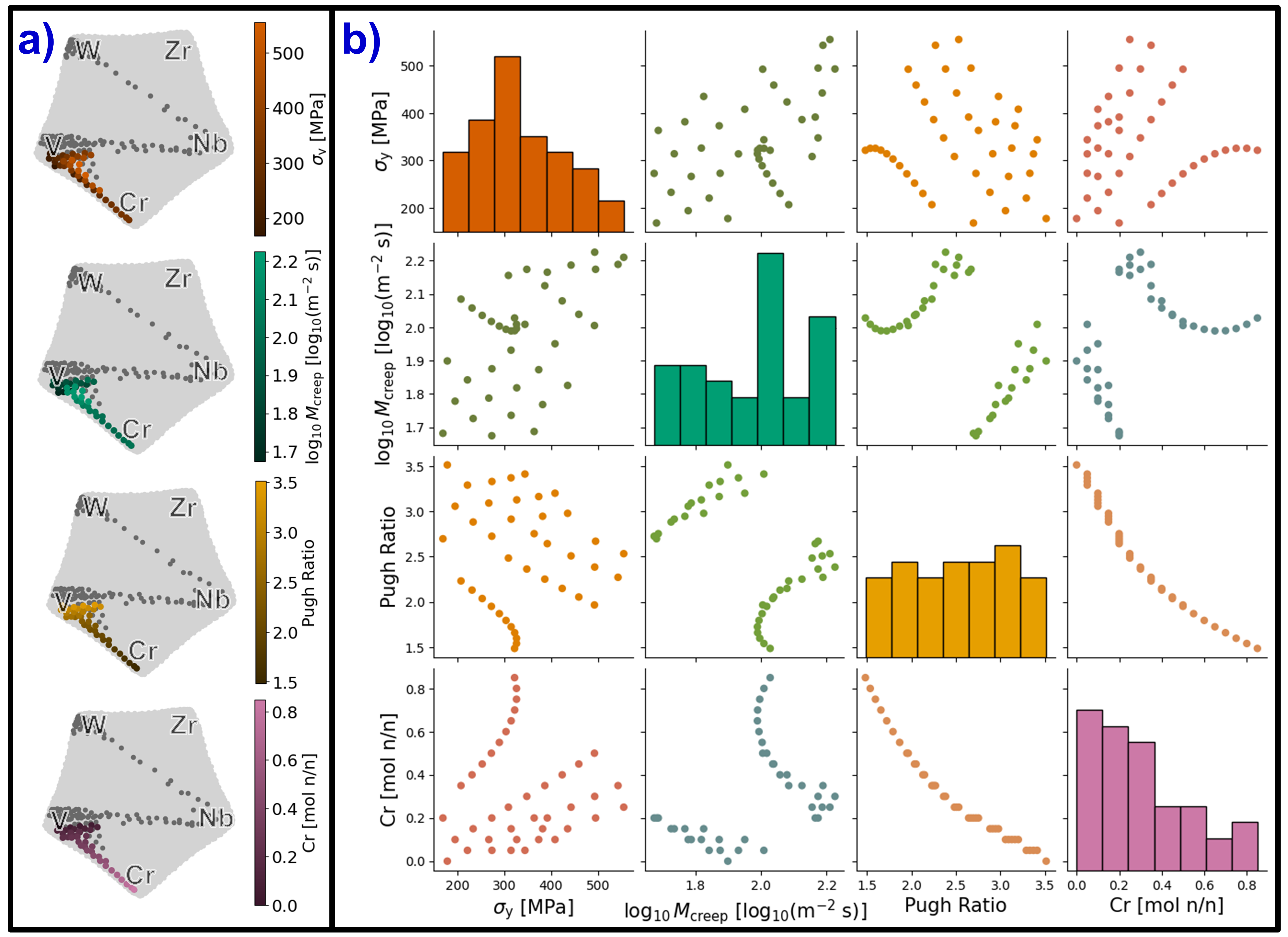}
\caption{\textbf{Relevant material properties for the $Alloy$ nodal compositions in the second-largest constrained subgraph $G_{\beta}$. a)} multiple color maps illustrating the qualitative variation of the properties of interest with respect to alloy composition using Uniform Manifold Approximation and Projection (UMAP) embedded coordinates, and \textbf{b)} Scatterplot matrix of the properties of interest for this work illustrating the pairwise correlation between properties (off-diagonal) and the distribution of each property (diagonal) for the reachable nodal compositions.}\label{fig:subgraph_2_space_analysis}
\end{figure}

For the second-largest subgraph, $G_{\beta} \subset G_{\text{AM}}$, color maps for the UMAP embedded coordinates, shown in Fig.~\ref{fig:subgraph_2_space_analysis}a, were used to qualitatively illustrate the properties of interest in relation to alloy composition. Similarly to $G_{\alpha}$, we saw that $\sigma_\text{y}$ and $M_{\mathrm{creep}}$ appear to decrease for compositions with higher V or Cr content and less configurational entropy. Meanwhile, the Pugh ratio seems to have a negative correlation with Cr content. Lastly, the color map of Cr content indicates the presence of the alloy compositions with up to the highest Cr content in the overall satisfactory alloy design space, 0.85 mole fraction. The scatter plot matrix for the nodal alloy compositions in $G_{\beta}$, shown in Fig.~\ref{fig:subgraph_2_space_analysis}b, indicates that almost all of the compositions have below average values for $\sigma_\text{y}$, with none exceeding the 95th percentile threshold of 771 MPa required for the turbine blade interior and base materials. Thus, despite the availability of alloys with higher Cr content and $M_{\mathrm{creep}}$, we did not select terminal alloy compositions from $G_{\beta}$ due to the low $\sigma_\text{y}$ values of its alloy compositions. This ultimately left $G_{\alpha}$ as the only viable subgraph for selecting terminal alloy compositions based on our property requirements.

\subsection{Multi-terminal CGA Design}

A general multi-terminal formulation of the CGA design problem \cite{allenGraphAlgorithmDesign2024, allenSystematicDesignCompositionally2025} can be defined as a minimum Steiner tree problem in graphs (STP) \cite{dreyfusSteinerProblemGraphs1971}. The STP is a generalization of two other classic optimization problems of 1) shortest paths and 2) minimum spanning trees, and can be described as follows: given an undirected graph $G = (V, E)$, a set of terminal nodes $V_{T} \subseteq V$, and a positive cost function $c(v_{j},v_{k})~ \forall~(v_{j},v_{k}) \in E$, find the tree $\tau_{\text{min}} = (V_\tau,E_\tau)$ which contains all terminal nodes such that $V_{T} \in V_\tau$ with minimal overall cost on $E_\tau$.

With this formulation, the terminal Steiner nodes are the terminal $Alloy$ nodal compositions in an ATLAS materials graph, which has been constrained for feasibility. An existing heuristic solver \cite{mehlhornFasterApproximationAlgorithm1988a} can be used to approximate the optimal tree, which minimizes a cost/objective function over the graph edges to optimize a compositional gradient between the given terminal alloys. The resultant CGA is represented as a graph tree, $\tau_{\text{CGA}} = \tau_{\text{min}}$. This approach enables the computational design of CGAs between any number of terminal alloys simultaneously, expanding computational CGA design beyond the one-dimensional paths of existing methods to multi-terminal material trajectories represented by graph trees, which can unify any number of terminal alloys.

To design a CGA between three terminal alloy compositions, the solver implementation by Mehlhorn for the minimum Steiner Tree Problem (STP) in weighted graphs was used \cite{mehlhornFasterApproximationAlgorithm1988a}, specifically as implemented in NetworkX \cite{hagbergExploringNetworkStructure2008} to design a multi-terminal CGA tree between the $Alloy$ nodes corresponding to the three terminal alloys. The solution given by this solver is guaranteed to have a total weight/cost no more than $2 \left( 1 - \frac{1}{l} \right)$ times that of a Steiner minimal tree, where $l$ is the minimum number of leaves in the optimal tree. Since there are three terminal nodes, the worst case is that there are three leaves, making the cost of the approximated Steiner minimal tree, at most, $\frac{4}{3}$ times that of a true Steiner minimal tree.

\subsection{Conformal CGA Mapping} 

The Tree-based Material Adjacency Propagation (TreeMAP) algorithm enables conformal mapping of an arbitrary multi-terminal gradient from a graph tree, $\tau_{\text{CGA}}$, onto a corresponding discretized 3D structural model \cite{allenGraphAlgorithmDesign2024, allenSystematicDesignCompositionally2025}. As input, TreeMAP takes a graph representation of the discretized part for rapid traversal, a gradient tree $\tau_{\text{CGA}}$, and a so-called \lq{coalescent material}\rq~parameter $m_c$, which is a user-specified node from $\tau_{\text{CGA}}$. Once complete, TreeMAP returns the part graph representation fully labeled with the gradient, which can then be mapped back to the discretized structural model.

\subsubsection{3D Structural Model Processing}

In Python, a 3D model was imported as a mesh from an STL file of a gas turbine blade publicly available from GrabCAD \cite{grabcad_gas_turbine_blade}. This surface mesh was converted into a voxelization with axial step sizes of $(\text{d}x, \text{d}y, \text{d}z) = (0.4, 0.4, 1.0)$ in millimeters using the \textit{pyvista.voxelize} function from PyVista. These step sizes were based on the reported hatch spacing and the compositional resolution in the build direction for multi-material AM with steels using LP-DED by Salas et al. \cite{salasEmbeddingHiddenInformation2022a}. Here, the $x$- and $y$-axis resolutions were based on the reported hatch spacing, and the $z$-resolution was based on the reported number of layers in the build direction to achieve a distinct compositional band. From this discretized representation, we generated an intermediate graph representation using NetworkX \cite{hagbergExploringNetworkStructure2008} as input for TreeMAP by defining a node with an index corresponding to each voxel, with edges indicating physically adjacent voxel pairs, resulting in a so-called part graph $G_P = (V_P, E_P)$.

\subsubsection{Initial Terminal Alloy Placement}
The terminal alloy for the exterior surface of the turbine blade, $x_{A}$ was located and placed within the initial part geometry by using PyVista to import a mesh file encompassing the blade region, which translates to $z > -3$ in the STL mesh file coordinates. This mesh was used to select the voxels in the entire blade $z > -3$, which we then down-selected to the surface voxels of this selection. These surface voxels were labeled with $Material\_ID = 1474$ for $x_{A}$.

The terminal alloy for the base of the turbine blade, $x_{B}$, was located based on the location of $x_{A}$ and the number of edges (16) in the tree between $x_{A}$ \& $x_{B}$ in $\tau_{\text{CGA}}$ to ensure the exact spacing for placing the pairwise gradient is achieved. All voxels more than 16 edges away from $x_{A}$ in the part graph $G_{P}$, and with $z < -3$ (to avoid encroaching on the blade core), were assigned $x_{B}$ with $Material\_ID = 7191$.

The placement of the terminal alloy $x_{C}$ could be determined similarly to $x_{B}$, since it is dependent on the location of the other terminal alloys. However, in this work, we defined the TreeMAP parameter $m_{c} = x_{C}$ ($Material\_ID = 2840$). This parameter defines the composition where the gradients propagating from each leaf node in $\tau_{\text{CGA}}$ \lq{coalesce}\rq, marking the final material propagation step in the TreeMAP algorithm. After the necessary compositional gradient steps are placed, this composition is used to fill the remaining unlabeled space. This is due to the assumption that the gradient alloys are not necessarily optimized for performance like the terminal alloys are; thus, they are placed in the minimum discrete steps, and any remaining space can be filled with an alloy with desired performance, given as the parameter $m_{c}$. In future work, we plan to further investigate the optimal distribution of CGAs.

\subsubsection{Conformal CGA Placement}
The CGA was placed using the TreeMAP algorithm \cite{allenGraphAlgorithmDesign2024, allenSystematicDesignCompositionally2025}. As input, we passed the gradient $\tau_{\text{CGA}}$, the spatial part graph $G_{P}$ labeled with the initial terminal alloys, and the parameter $m_{c} = 2840$. After running TreeMAP, the nodes $V_P \in G_{P}$ were all labeled with a material identifier corresponding to $\tau_{\text{CGA}} \in G_{\text{AM}}$ at each part graph node $v_{P,i} \in V_P$. Mapping these labels from $V_P$ back to the corresponding voxels results in a volumetric structural representation with completely defined CGA placement.

\section*{Data Availability}

All primary data relevant to the findings of this work have been included in the main text. The materials dataset supporting the findings of this study is available at Zenodo \cite{allen2024nbcrvwzr} and can be accessed via the DOI: \href{https://doi.org/10.5281/zenodo.14220614}{10.5281/zenodo.14220614}.

\section*{Code Availability}

No new custom code was used to generate the central results in this work. For more information about ATLAS and TreeMAP, we refer readers to our prior work \cite{allenGraphDatabaseSchema2024, allenGraphAlgorithmDesign2024, allenSystematicDesignCompositionally2025}


\bibliography{sn-article}


\backmatter

\section*{Acknowledgments}
We acknowledge grant number NSF-DGE-1545403 (\emph{NSF-NRT: Data-Enabled Discovery and Design of Energy Materials, D$^3$EM}). MA and RA wish to acknowledge NSF through Grant No. NSF-DMREF-2323611. MA and RM acknowledge partial support from the J. Mike Walker '66 Department of Mechanical Engineering. VA was partially supported by ARL under Contract No. W911NF-22-2-0106. VA and RA also acknowledge the support from the Army Research Office through Grant No. W911NF-22-2-0117. BV acknowledges the support of NSF through Grant No. 1746932. Portions of this research were conducted with the advanced computing resources provided by Texas A\&M High-Performance Research Computing.

\section*{Author Contributions}

M.A. collected the property and creep data, developed the design methods, performed the application of these methods, and led the analysis and writing of the manuscript. V.A. developed and trained the machine learning model used in the study, and wrote the initial draft of the Deep Learning Property Regression Model sections. B.V. collected and curated the phase datasets, wrote the initial draft of the Thermodynamic Calculations section, and contributed to the property and creep modeling scripts used for data prediction. J.H. wrote the initial draft of the Material Property Modeling section and contributed to the property and creep modeling scripts used for data prediction. R.M. \& R.A. contributed through funding acquisition, project administration, resources, supervision, and validation. All sections were edited and finalized by M.A., and all authors reviewed and provided feedback on the manuscript.

\section*{Competing Interests}

The authors declare no competing interests.



\begin{appendices}

\section{Deep Learning Property Regression Model}\label{secA1}

The encoder's architecture consists of three layers with neuron counts of 256, 512, and 1024, each layer utilizing LeakyReLU activations (with \(\alpha = 0.0164\)), batch normalization, L2 regularization (\(\lambda = 1 \times 10^{-5}\)), and a dropout rate of 0.2 to mitigate overfitting and enhance model generalizability. This overcomplete model, with a latent dimension of 2048 that exceeds the input dimension, allows for a richer representation of complex patterns in the data. The decoder mirrors the encoder's structure with three layers, reducing the dimensionality from the latent space back to the original feature space through neuron counts of 1024, 512, and 256. This architecture enables the model to effectively learn mappings from input features to predicted properties, even in high-dimensional data environments. The model was trained by minimizing the Mean Squared Error (MSE) between predicted and true values, optimizing weights to improve prediction accuracy while leveraging the flexibility of the overcomplete latent space~\cite{attari2024decoding}.

For training, the dataset was divided into 80\% for training, 10\% for validation, and 10\% for testing. This split ensured a robust evaluation of model performance. Key evaluation metrics, including mean squared logarithmic error (MSLE), root mean squared logarithmic error (RMSLE), log-transformed coefficient of determination (\(\log R^2\)), geometric mean absolute error (GMAE), symmetric mean absolute percentage error (SMAPE), mean absolute scaled error (MASE), and root mean squared percentage error (RMSPE), demonstrated reliable predictive accuracy for the properties of interest. This deep neural network architecture enables precise property predictions across diverse compositions, handling features with high skewness and kurtosis. The model's primary role in this study was to provide accurate predictions for necessary data points, supporting gradient-based design by contributing to the cost function calculation. For further details on the model architecture and its theoretical basis, we refer readers to our prior work~\cite{attari2024decoding}.

Figure~\ref{fig:fast_screening_properties} illustrates this model’s performance for various properties using parity plots (Fig.\ref{fig:fast_screening_properties}a-e) and a probability plot (Fig.\ref{fig:fast_screening_properties}f). The parity plots (a-e) show predicted versus actual values for properties such as yield strength at 1000$^\circ$C, Kou Criteria, minimum Coble creep rate \cite{cobleCreep1963} at 1000$^\circ$C, Pugh Ratio, and Creep Merit. The data points generally cluster close to the ideal line, demonstrating the model’s predictive accuracy. Figs.~\ref{fig:fast_screening_properties}(c) highlights challenges in predicting extreme creep values, where the data slightly diverges from the ideal line due to high skewness and kurtosis (calculated as \textbf{39} and 1.5$\times10^{3}$, respectively). Despite this, the model captures the overall trend, and the smaller insets demonstrate that quantile-transformed, normalized data improves the predictive performance. 

Figure~\ref{fig:fast_screening_properties}(f) provides further validation through a quantile-quantile (QQ) plot comparing theoretical quantiles to ordered values of all output features. The true and predicted data points align closely. The ideal fit line (normal distribution) demonstrates that the dataset deviates from normality, further emphasizing the model’s effectiveness in capturing the underlying data distribution. The inset in Fig.~\ref{fig:fast_screening_properties}(f) shows the training and validation loss history, illustrating steady convergence. The model’s ability to quickly predict material properties in unexplored chemistry spaces enabled efficient alloy design with desired properties while significantly reducing computational costs compared to traditional methods.




\end{appendices}


\end{document}